\documentclass[a4paper,11pt]{article}

\usepackage[hmargin=25mm,vmargin=25mm]{geometry}
\usepackage[T1]{fontenc}
\usepackage{lmodern}
\usepackage{microtype}

\usepackage{enumitem}
\usepackage{amsmath,amssymb,mathtools,bm,amsthm}
\usepackage{mathtools,bm}
\usepackage{pifont}
\usepackage{graphicx,booktabs,array,tabularx,multirow}
\usepackage{algorithm,algorithmic}
\usepackage{tikz}
\usetikzlibrary{arrows.meta,positioning,calc}

\usepackage[authoryear]{natbib}
\setcitestyle{square,citesep={,},aysep={}}
\usepackage[hidelinks]{hyperref}
\usepackage[nameinlink,capitalize,noabbrev]{cleveref}

\theoremstyle{plain}
\newtheorem{theorem}{Theorem}
\newtheorem*{maintheorem}{Main Theorem}
\newtheorem{lemma}[theorem]{Lemma}

\newtheorem{proposition}[theorem]{Proposition}

\newtheorem{remark}[theorem]{Remark}

\crefname{section}{Section}{Sections}
\Crefname{section}{Section}{Sections}
\crefname{equation}{Equation}{Equations}
\Crefname{equation}{Equation}{Equations}
\crefname{appendix}{Appendix}{Appendices}
\Crefname{appendix}{Appendix}{Appendices}

\newcommand{\R}{\mathbb{R}}
\newcommand{\Z}{\mathbb{Z}}
\newcommand{\cB}{\mathcal{B}}
\newcommand{\cC}{\mathcal{C}}

\newcommand{\cF}{\mathcal{F}}

\newcommand{\cBZ}{\cB_{\Z}}
\newcommand{\LW}{\operatorname{LW}}
\newcommand{\SW}{\operatorname{SW}}
\newcommand{\LWO}{\LW^{\mathrm{OPT}}}
\newcommand{\Rev}{\operatorname{Rev}}
\newcommand{\xf}{\bm{x}^{\mathrm f}}
\newcommand{\yf}{\bm{y}^{\mathrm f}}
\newcommand{\zf}{\bm{z}^{\mathrm f}}
\newcommand{\pf}{\bm{p}^{\mathrm f}}
\newcommand{\eps}{\varepsilon}
\newcommand{\ind}{\mathbf{1}}
\newcommand{\unit}{\bm{\chi}}

\renewcommand{\mid}{\mathrel{:}}

\newif\ifwithappendices
\ifdefined\SubmissionVersion
  \withappendicesfalse
  \expandafter\let\csname r@TotPages\endcsname\relax
\else
  \withappendicestrue
\fi

\title{Clinching Auctions for Mixed Divisible and Indivisible~Goods}
\author{Ryosuke Sato\thanks{School of Science and Engineering, Chuo University, Tokyo, Japan. \texttt{rsato636@g.chuo-u.ac.jp}}}

\begin{document}
\maketitle

\begin{abstract}
We consider budget-constrained auctions for homogeneous mixed goods
with public budgets, where divisible and indivisible goods are jointly
allocated and each buyer has the same per-unit value for both.
This setting captures markets in which equal-sized lots must be allocated
intact, while the same commodity is available to all buyers as divisible stock.
We extend the celebrated clinching framework to this setting with an auction that determines total quantities and payments as the price rises, while deferring the decomposition of these quantities into divisible and indivisible allocations until termination.
Our deterministic mechanism satisfies incentive compatibility,
individual rationality, budget feasibility, and Pareto optimality.
It also achieves a 1/2-approximation to optimal liquid welfare. This is the best possible guarantee over mixed-goods instances among deterministic mechanisms satisfying incentive compatibility and budget feasibility.
These guarantees hold under arbitrary integer polymatroid constraints on indivisible allocations and a common divisible supply. We further show that restricting a single buyer's access to the divisible supply can make incentive compatibility, individual rationality, and Pareto optimality incompatible for deterministic mechanisms with budget-feasible, nonnegative payments, even with three buyers.
\end{abstract}

\section{Introduction}\label[section]{sec:introduction}
Budget-constrained auctions seek to allocate goods efficiently while
inducing truthful bidding and respecting buyers' spending limits.
In advertising markets, for example, an advertiser may value additional
impressions highly but have little budget left.
These budget constraints make it difficult to induce truthful bidding
while maximizing social welfare (SW), the total value of the allocation.
Even with public budgets, no budget-feasible mechanism with
nonnegative payments can guarantee a constant approximation to optimal
SW while satisfying incentive compatibility (IC) and individual
rationality (IR)~\citep{DobzinskiLaviNisan2012,DobzinskiPaesLeme2014}.

\citet{DobzinskiLaviNisan2012} achieved IC, IR, and Pareto optimality (PO) for budget-constrained auctions using a variant of Ausubel's clinching auction~\citep{A2004}.
Clinching auctions allocate goods at the current price during an ascending-price process while preserving other buyers' feasible allocations.
Under polymatroid constraints, \citet{GMP2015} developed a polyhedral clinching auction for divisible goods, and \citet{HiraiSato2025} constructed its indivisible counterpart.
Both auctions satisfy IC, IR, and PO with public budgets.
Beyond these properties, liquid welfare (LW), which caps each buyer's value contribution at her budget, provides an efficiency benchmark for budget-constrained auctions.
The $1/2$-approximation to optimal LW, originally established for unconstrained divisible goods~\citep{DobzinskiPaesLeme2014}, also holds for polyhedral clinching auctions with either type of goods~\citep{Sato2026,HiraiSato2025}.

These results treat divisible and indivisible goods separately.
In practice, the same commodity may be sold in equal-sized lots,
each assigned intact to one buyer, and as stock that can be divided freely.
For example, a seller may offer the same oil in sealed containers
of a fixed volume and in any quantity from a tank.
Similarly, a publisher may offer equal-sized packages of ad impressions,
each sold whole, alongside individual impressions for the same audience.
At the scale of a package, the latter supply can be treated as divisible.
In either case, when a buyer has the same per-unit valuation for both forms,
her value depends only on the total quantity she receives.
Feasibility still depends on how that quantity is split between
equal-sized lots and divisible stock.

Extending clinching auctions to such mixed goods is not straightforward.
Running the divisible and indivisible auctions separately requires
splitting each buyer's budget between them. If all goods are instead
treated as divisible, the total quantities at termination may admit
no feasible decomposition into divisible and indivisible goods.
These challenges raise the following~questions:
\begin{itemize}
\item Can a single auction allocate mixed goods while satisfying IC, IR, and PO?
\item Can it retain the $1/2$-approximation to optimal LW?
\item What allocation constraints can it accommodate?
\end{itemize}
This paper addresses all of the questions by developing a unified clinching auction for mixed~goods. 
\begin{table*}[t]
\centering
\small
\begin{tabularx}{\textwidth}{@{}>{\raggedright\arraybackslash}X l l c c c@{}}
\toprule
\multirow{2}{*}{Result}
& \multicolumn{2}{c}{Constraints}
& \multirow{2}{*}{PO}
& \multicolumn{2}{c}{LW bounds}\\
\cmidrule(lr){2-3}\cmidrule(l){5-6}
& Indivisible & Divisible & & Guarantee & Upper\\
\midrule
\citet{BCMX2010} \& \citet{DobzinskiPaesLeme2014}
& --- & Common supply & \ding{51} & $1/2$ & $3/4$\\
\citet{GMP2015} \& \citet{Sato2026}
& --- & Polymatroid & \ding{51} & $1/2$\textsuperscript{*} & $3/4$\\
\citet{HiraiSato2025}
& Integer polymatroid & --- & \ding{51} & $1/2$ & $3/4$\\
\addlinespace[2pt]
\textbf{This study} (Main Theorem \& \Cref{thm:lower-bound})
& Integer polymatroid & Common supply & \ding{51} & $1/2$ & $1/2$\\
\textbf{This study} (\Cref{thm:eligibility-impossible})
& One unrestricted item & Buyers 1--2 only
& \ding{55}\textsuperscript{\(\dagger\)} & --- & ---\\
\bottomrule
\end{tabularx}
\caption{PO and LW guarantees for clinching auctions.
Upper bounds apply to all deterministic budget-feasible mechanisms that satisfy IC.
Dashes under Constraints indicate absent good types.
\textsuperscript{*}Under an assumption on the price increment.
\textsuperscript{\(\dagger\)}No deterministic mechanism with budget-feasible,
nonnegative payments satisfies IC, IR, and PO on this instance.}
\label[table]{tab:comparison}
\end{table*}

\subsection{Our Contributions}
We study auctions for homogeneous mixed goods under public budgets.
Indivisible allocations are subject to an arbitrary integer polymatroid constraint, while a common divisible supply is available to all buyers.
Our auction recovers the clinching auctions for purely divisible
goods~\citep{BCMX2010} and purely indivisible goods~\citep{HiraiSato2025}.

We develop a deterministic clinching auction that fixes total quantities and payments as the price rises and decomposes the quantities into indivisible and divisible goods only at termination.
To see why we defer the decomposition, consider a buyer who raises her bid and can therefore stay active longer.
If each purchase fixed the type of good, she would keep every divisible unit she would have received at the lower bid.
Even with equal budgets and one indivisible item, however, IC, IR, and PO with nonnegative payments can require her to receive fewer divisible units at the higher bid (\Cref{prop:fluiddecrease}).

As in the pure settings, each buyer clinches at the current price
any additional quantity she can receive without restricting the
other buyers' feasible allocations. Our auction also uses this price
and the remaining budgets to update an upper bound on each buyer's
indivisible allocation. Tightening one buyer's bound can cause another
to clinch immediately. We show that the quantities clinched so far remain consistent with a
feasible final allocation, even when the sums of feasible divisible
and indivisible allocations do not form a polymatroid. 
Our main result is summarized as follows:
\begin{maintheorem}
The proposed auction (\Cref{alg:main}) is budget feasible and satisfies IC, IR, and PO.
Moreover, it achieves a $1/2$-approximation to optimal LW.
\end{maintheorem}
Notably, our LW guarantee is best possible among deterministic, budget-feasible IC mechanisms over mixed-goods markets (\Cref{thm:lower-bound}).
This contrasts with the purely divisible and indivisible settings, where a gap remains between the $1/2$-approximation of clinching auctions and the $3/4$ upper bound~\citep{DobzinskiPaesLeme2014,HiraiSato2025}.
\Cref{tab:comparison} summarizes these results.
Of independent interest, our proof of the LW guarantee is much shorter than that of \citet{HiraiSato2025}, while also covering mixed goods.

Despite the impossibility of a constant SW approximation, we also show that mechanisms satisfying IR and PO achieve SW of at least optimal LW (\Cref{prop:po-sw}).
This extends the SW guarantee previously established for polyhedral clinching auctions in the purely divisible and purely indivisible settings~\citep{Sato2026,HiraiSato2025}.

We also show that restricting one buyer's access to divisible goods
can make IC, IR, and PO incompatible for deterministic mechanisms
with budget-feasible, nonnegative payments, even with three buyers
(\Cref{thm:eligibility-impossible}).
For every $0<\eta<1$, such an instance exists even when two buyers
have unrestricted access and the third can receive up to a fraction
$\eta$ of the divisible supply.
\Cref{app:constraint-transfer} gives sufficient conditions for
implementing divisible clinching outcomes under a common polymatroid
constraint on total quantities or separate constraints on the two
types of goods, while preserving the main theorem's guarantees.

\subsection{Related Work}
Beyond homogeneous multi-unit auctions~\citep{DobzinskiLaviNisan2012}, clinching-based mechanisms have been developed for matching markets~\citep{FLSS2011} and polymatroidal environments~\citep{GMP2015}. While \citet{DuettingHenzingerStarnberger2015} prove impossibility
results for budget-constrained auctions with heterogeneous goods,
our impossibility arises from mixed divisibility and restricted
access to divisible goods, even with equal per-unit values across types.
\citet{ColiniBaldeschiEtAl2015} randomly round a clinching auction's
outcome for divisible ad slots to an integral allocation.
Their IC and IR guarantees hold in expectation,
whereas our mechanism achieves both deterministically.
Clinching auctions have recently drawn interest in credible
auction design~\citep{Saitto2026}.

LW was independently introduced as an efficiency benchmark for budget-constrained auctions by \citet{DobzinskiPaesLeme2014} and \citet{SyrgkanisTardos2013}. It has since become a widely used benchmark beyond clinching auctions (e.g.,~\citep{FT2023,FLS2024}). \citet{LuXiao2015} improve upon the $1/2$ approximation of clinching auctions for homogeneous divisible goods, whereas our $1/2$ guarantee is best possible across mixed-goods markets.
 
Mixed-goods allocation has attracted growing interest, especially
in fair division~\citep{BeiEtAl2021,LiEtAl2023}.
For a survey, see \citet{LiuEtAl2024}.
\citet{KNS2026} study the Minkowski sum of an integral base
polyhedron and the integer points of another, with applications
to fair allocation.
A similar structure arises in our model under separate polymatroid
constraints, whereas our focus is efficiency under IC and public~budgets.

\subsection{Notation}
We write $\Z_+=\{0,1,\ldots\}$ for the nonnegative integers.
For a vector $\bm{w}$ indexed by a finite set $N$ and any $S\subseteq N$,
let $\bm{w}(S)=\sum_{i\in S}w_i$.
In set expressions, we abbreviate the singleton $\{i\}$ by $i$.
We denote the $i$-th unit vector by $\unit_i$ and the positive part of
$t\in\R$ by $[t]_+=\max\{t,0\}$.
For vectors, we interpret inequalities, $\min$, and $\lfloor\cdot\rfloor$ componentwise.

\section{Our Model}\label[section]{sec:model}
A seller allocates $l\in\Z_+$ indivisible units and $m\ge0$
units of a divisible good to buyers in $N\coloneqq\{1,\ldots,n\}$\ ($n\geq 2$),
with total supply $s\coloneqq l+m>0$.
Each buyer $i$ has a private per-unit valuation $v_i>0$, common to
both types, and reports a bid $v'_i>0$.
Her public budget $B_i>0$ limits her payment.

An \emph{allocation} $(\bm{y},\bm{z})$ gives buyer $i$
an integer quantity $y_i$ of indivisible goods and a divisible
quantity $z_i$.
Her total quantity is $x_i=y_i+z_i$, with valuation $v_i x_i$.
Indivisible allocations satisfy a polymatroid constraint defined by a
monotone submodular function $f\colon 2^N\to\Z_+$ with $f(\varnothing)=0$.
Monotonicity means that $f(S)\le f(T)$ whenever $S\subseteq T\subseteq N$,
and submodularity means that $f(S)+f(T)\ge f(S\cup T)+f(S\cap T)$ for all
$S,T\subseteq N$.
We also assume that $f(N)=f(N\setminus i)=l$ for every buyer $i$.
Thus, all indivisible units are initially open to competition, since
the full supply can be allocated even if any one buyer is excluded.\footnote{For general $f$, first allocate \,$l-f(N\setminus i)$\,\, indivisible units
to each buyer $i$ for free. The residual instance satisfies this
assumption, and the reduction preserves all guarantees
(see \Cref{app:mandatory-quantities}).} 
For example, setting $f(S)=l$ for every nonempty $S$ allows any allocation of up to $l$ indivisible units.
Other choices of $f$ can limit the number of units assigned to individual buyers or the total assigned to each of several disjoint buyer groups. 
Divisible allocations are subject only to the total supply limit $m$.
An allocation is called \emph{feasible} if
$\bm{y}\in P(f)\cap\Z_+^N$,\ 
$\bm{z}\ge0$,\ $\bm{z}(N)\le m$, 
where $P(f)=\{\bm{y}\ge0\mid\bm{y}(S)\le f(S)\ (S\subseteq N)\}$.
Let $\cF$ be the set of all feasible allocations. 

A \emph{mechanism} $\mathcal M$ maps bids $\bm{v}'$ to an
\emph{outcome} $(\bm{y},\bm{z},\bm{p})$, consisting of a feasible
allocation and payments.
It is \emph{budget feasible} if every output satisfies $p_i\le B_i$,
and makes \emph{no positive transfers} if $p_i\ge0$.
Buyer~$i$'s utility is $u_i(\bm{v}';\mathcal M)=v_i x_i-p_i$ if
$p_i\le B_i$ and $-\infty$ otherwise.
The seller's utility is the revenue $\Rev\coloneqq\bm{p}(N)$.
For every buyer $i$ and all valuation and bid profiles, we require both:
\begin{itemize}[leftmargin=*]
\item Incentive Compatibility (IC).
Truthful bidding maximizes utility:
$u_i((v_i,\bm{v}'_{-i});\mathcal M)
\ge u_i(\bm{v}';\mathcal M)$.
\item Individual Rationality (IR).
Truthful bidding yields nonnegative utility:
$u_i((v_i,\bm{v}'_{-i});\mathcal M)\ge~0$.
\end{itemize}
Here $\bm{v}'_{-i}$ denotes the vector of other buyers' bids.

An outcome $(\bm{y},\bm{z},\bm{p})$ is \emph{Pareto optimal}
if no alternative allocation $(\bm{y}',\bm{z}')\in\cF$
and payments $\bm{p}'\le\bm{B}$ satisfy
$v_i(y_i'+z_i')-p_i' \ge v_i x_i-p_i \ (i\in N)\ \ {\rm and}\ \   
\bm{p}'(N)\ge\bm{p}(N)$,
with at least one strict inequality.
Comparisons allow negative payments, while our auction's payments are nonnegative.
A mechanism satisfies \emph{Pareto optimality (PO)} if truthful bidding
yields a Pareto optimal outcome for every valuation profile.

We evaluate efficiency at buyers' true valuations.
For $\bm{x}=\bm{y}+\bm{z}$,
\emph{social welfare (SW)} sums buyers' valuations for their allocations,
while \emph{liquid welfare (LW)} caps each buyer's contribution at her budget.
The measures and benchmark are
\begin{align}\label{eq:welfare-definition}
\SW(\bm{x})=\sum_i v_i x_i,\notag \quad
\LW(\bm{x})=\sum_i\min\{v_i x_i,B_i\},\notag \quad
\ \LWO=\max_{(\bm{y},\bm{z})\in\cF}\LW(\bm{y}+\bm{z}).
\end{align}

The following proposition shows that the SW guarantee for divisible and indivisible clinching auctions~\citep{Sato2026,HiraiSato2025} follows from IR and PO alone.
\begin{proposition}\label[proposition]{prop:po-sw}
Let $\bm{x}$ and $\bm{p}$ be the total allocation and payments
of a budget-feasible, Pareto optimal outcome with nonnegative utility for every buyer.
For every feasible total allocation $\bm{x}'$,
$\SW(\bm{x})\ge\LW(\bm{x}')$. Hence, $\SW(\bm{x})\ge\LWO$.
\end{proposition}
\begin{proof}
Let $u_i\coloneqq v_i x_i-p_i\ge0$ denote buyer $i$'s utility.
Choose payments $p_i'=\min\{v_i x_i',B_i\}-u_i$.
Then, $p_i'\le B_i$, and each buyer's utility satisfies
$v_i x_i'-p_i'
=u_i+\max\{v_i x_i'-B_i,0\}\ge u_i$.
Since the outcome is Pareto optimal and no buyer is worse off,
the seller's revenue cannot increase. Hence, 
$0\ge\bm{p}'(N)-\bm{p}(N)=\LW(\bm{x}')-\SW(\bm{x})$.
\end{proof}
This guarantee is relative to optimal LW.
With nonnegative payments, no budget-feasible mechanism satisfying IC and IR can guarantee a constant-factor approximation to optimal SW~\citep{DobzinskiPaesLeme2014}.
We focus on PO and LW approximation in what follows.

\subsection{Application}
Guaranteed advertising contracts specify quantities of impressions,
while spot markets sell impressions individually~\citep{GhoshEtAl2009}.
Consider a publisher that sells guaranteed impressions in equal-sized blocks, each assigned intact to one advertiser.
The publisher also sells reserve impressions from the same audience segment in any quantity, and each is delivered with a publicly known probability $q\in(0,1]$.
For an advertiser with constant expected value per delivered impression, a reserve impression is worth $q$ times as much as a guaranteed one.
We count each reserve impression as $q$ guaranteed impressions, measure quantities in block units, and treat the reserve as divisible.
Each advertiser has a single public budget for purchases from both sources.
Limits on the number of blocks assigned to individual advertisers or disjoint groups define integer polymatroid constraints.
Supply and budgets are fixed before bidding, and ads are shown
after the~auction.

\subsection{Unequal Per-Unit Values}
The same rescaling applies whenever the ratio $q>0$ of divisible to indivisible per-unit values is public and common to all buyers.
With $\tilde z_i=qz_i$, buyer~$i$'s value is $v_i(y_i+\tilde z_i)$, and our guarantees hold with divisible supply $qm$.
If the ratios $q_i$ are public but differ across buyers, setting $\tilde z_i=q_i z_i$ gives the supply constraint $\sum_i\tilde z_i/q_i\le m$.
For $m>0$, this region is not a polymatroid, and the guarantees of \citet{GMP2015} therefore do not apply directly.
If $q_i$ is private, buyer~$i$ has a two-dimensional type $(v_i,q_i)$.
\citet{DuettingHenzingerStarnberger2015} prove impossibility results in the related setting of heterogeneous goods.
All our guarantees thus extend to unequal per-unit values across the two forms whenever buyers share a publicly known ratio.

\section{The Mixed Clinching Auction}\label[section]{sec:auction}
We first explain why our mechanism defers the decomposition until termination.
We then describe its update rules and establish feasibility, incentive
compatibility, and individual rationality.

\subsection{Why We Defer the Decomposition}
As the price rises, our auction determines each buyer's total quantity
and payment. Only at termination does it split each buyer's total
between indivisible and divisible goods.
We show why fixing the type of good at each purchase can conflict
with IC, IR, and PO.

Consider a clock auction in which each report is used only as the buyer's dropping price.
Suppose that the auction \emph{commits allocations as the clock proceeds}: goods once purchased are never taken back or exchanged for other goods, and no purchases occur at or after the buyer's dropping.
Such an auction makes the divisible quantity assigned to each buyer weakly increasing in her own report.
Fix a buyer $i$ and consider increasing her report from $a$ to $b>a$, holding all other reports $\bm v_{-i}$ fixed.
Let $z_i(\bm v)$ denote the quantity of divisible goods assigned to buyer $i$ at termination under report profile $\bm v$.
Both runs allocate the same divisible quantity to buyer $i$ before price $a$.
Since she drops out at $a$ under the lower report and this quantity is committed under the higher report, we obtain $z_i(b,\bm v_{-i})\ge z_i(a,\bm v_{-i})$.
However, IC, IR, and PO can force a buyer's divisible allocation to
fall by almost the entire supply when she raises her report.
\begin{proposition}\label[proposition]{prop:fluiddecrease}
For every divisible supply $m\in(0,1)$, there is an instance with the
following property. For any deterministic, budget-feasible mechanism
satisfying IC, IR, and PO with nonnegative payments and any
$\varepsilon\in(0,m)$, there exist a buyer $i$, fixed reports
$\bm v_{-i}$, and reports $a<b$ such that
$z_i(a,\bm v_{-i})-z_i(b,\bm v_{-i})>m-\varepsilon$.
\end{proposition}
The proof in \Cref{app:polimits} uses three buyers with the same public
budget $B>0$ and one unrestricted indivisible item.
The proposition shows that the clock rules above cannot guarantee
IC, IR, and PO. Our auction therefore fixes each buyer's total quantity
and payment as the price rises but defers the decomposition until
termination. A higher report can then lead to a smaller final divisible
allocation without undoing earlier purchases.

\subsection{The Auction}\label[section]{sec:auction-rules}
\Cref{alg:main} gives our auction, with the purchase and price updates defined below.
The auction raises a common price $c$.
At each price, an active buyer \emph{clinches} additional quantity
without restricting the other buyers' feasible allocations,
taking their remaining budgets and the allocation constraints into account.
Each unit is charged at the current price.
These purchases fix total quantities and payments,
while their decomposition into indivisible and divisible goods is determined at termination. We use the following notation:
\addtolength{\leftmargini}{-1em}
\begin{itemize}
\item $\bm{x},\bm{p}\in\R_+^N$ denote the buyers' current total quantities
and payments, respectively.
\item $A=\{i\in N\mid v'_i>c,\ p_i<B_i\}$ is the set of active buyers
between events.
\item $\bm{d}\in\Z_+^N$ gives the current upper bounds on buyers'
indivisible quantities. The final indivisible allocation $\bm{y}$
must satisfy $\bm{y}\le\bm{d}$.
\item $b_i=B_i-p_i\in[0,B_i]$ is the remaining budget of buyer~$i$.
\item $r=s-\bm{x}(N)\in[0,s]$ is the remaining supply of both types
of goods combined.
\end{itemize}
Initially, set $c=0$, $\bm{x}=\bm{p}=\bm{0}$, and $A=N$.
Since $f(N\setminus i)=l$ for every buyer~$i$, no initial allocation is needed.
All choices between buyers follow a fixed order independent of reports.
\begin{algorithm}[t]
\caption{Mixed clinching auction}\label[algorithm]{alg:main}
\begin{algorithmic}[1]
\REQUIRE Monotone submodular function $f$, divisible supply $m$, budgets $\bm{B}$, and bids $\bm{v}'$.
\STATE $c\gets0$, $\bm{x}\gets\bm{0}$, $\bm{p}\gets\bm{0}$, $A\gets N$.
\STATE $d_i\gets f(i)+1$ for all $i$.
\WHILE{$\bm{x}(N)<s$}
  \IF{some $i\in A$ satisfies $v'_i=c$}
    \STATE Choose the first such buyer $i$.
    \STATE $A\gets A\setminus i$, $d_i\gets\min\{d_i,\lfloor x_i\rfloor\}$.
    \STATE Compute $\delta_i$ for each $i\in A$ using \eqref{eq:clinch}.
    \STATE $(x_i,p_i)\gets(x_i+\delta_i,p_i+c\delta_i)$ for each $i\in A$.
  \ELSIF{some $i\in A$ satisfies $b_i=c(d_i-x_i)>0$}
    \STATE For the first such buyer $i$, set $d_i\gets d_i-1$.
    \STATE Compute $\delta_i$ for each $i\in A$ using \eqref{eq:clinch}.
    \STATE $(x_i,p_i)\gets(x_i+\delta_i,p_i+c\delta_i)$ for each $i\in A$.
  \ELSE
    \STATE Increase $c$ until the next event, updating quantities and budgets according to \eqref{eq:ode}.  \ENDIF
  \STATE For each $i\in A$ with $b_i=0$, set $d_i\gets\min\{d_i,\lfloor x_i\rfloor\}$ and remove $i$ from $A$.
\ENDWHILE
\STATE Greedily choose an integer base $\bm{y}\le\min\{\bm{d},\lfloor\bm{x}\rfloor\}$ in the fixed buyer order.
\RETURN $(\bm{y},\bm{z},\bm{p})=(\bm{y},\bm{x}-\bm{y},\bm{p})$.
\end{algorithmic}
\end{algorithm}

\subsubsection{Clinching at a Fixed Price (lines~7--8 and~11--12)}
For the bounds $\bm{d}$, define the rank function~\citep{GMP2015} by
$f_{\bm{d}}(S)=\min_{T\subseteq S}\{f(T)+\bm{d}(S\setminus T)\}$. 
Note that $P(f_{\bm{d}})=P(f)\cap\{\bm{y}\mid\bm{y}\le\bm{d}\}$.
If $f_{\bm{d}}(N)=l$, the minimum $i$-th coordinate among integer bases of $P(f_{\bm{d}})$ is
$\mu_i(\bm{d})=l-f_{\bm{d}}(N\setminus i)$. At price $c>0$, each active buyer $i$ buys
\begin{equation}\label{eq:clinch}
\delta_i=\max\{0,\ r-\bm{b}(A\setminus i)/c,\ \mu_i(\bm{d})-x_i\}
\end{equation}
units at price $c$, paying $c\delta_i$.
The budget term in \eqref{eq:clinch} is the clinching amount of the adaptive
clinching auction, and the integer term is that of \citet{HiraiSato2025} (\Cref{app:pure-equivalence}).
When another buyer purchases a quantity $q$, both $r$ and
$\bm{b}(A\setminus i)/c$ decrease by $q$, while $\bm{d}$ stays fixed.
Thus, buyer $i$'s clinching amount is independent of the purchase order.
We call the resulting state \emph{stable} if no active buyer can clinch
more at price $c$.

\subsubsection{Clinching as the Price Rises (line~14)}
At a stable state, define the \emph{clinching set}
$C\coloneqq\{i\in A\mid \bm{b}(A\setminus i)=cr\}$.
For each $i\in C$, the other active buyers' remaining budgets exactly cover the cost of the remaining supply.
If $C$ is empty, increase the price without purchases.
Otherwise, while $A$, $C$, and $\bm{d}$ remain fixed, quantities and budgets are updated by
\begin{align}\label{eq:ode}
\frac{dx_i}{dc}=\frac rc,\quad
\frac{db_i}{dc}=-r\ (i\in C),\qquad
\frac{dr}{dc}=-\frac{|C|r}{c}.
\end{align}
Outside $C$, buyers' quantities and budgets remain fixed.
The price increases until one of the following occurs:
an active buyer drops out, exhausts her budget, satisfies $b_i=c(d_i-x_i)>0$, or enters $C$.
A buyer entering $C$ begins clinching as the price rises, and \eqref{eq:ode} continues with the updated set.

\subsubsection{Termination and Final Allocation}
The auction stops when $\bm{x}(N)=s$, meaning that all supply has been purchased.
It then greedily selects an integer base
$\bm{y}\le\min\{\bm{d},\lfloor\bm{x}\rfloor\}$ in the fixed buyer order
and assigns the remaining quantities $\bm{z}=\bm{x}-\bm{y}$ as divisible goods.
Each buyer receives $x_i$ units in total, and her payment remains $p_i$. Section \ref{sec:auction-feasibility} establishes the feasibility of this final step.

\subsubsection{Example}
A decrease in one buyer's integer bound can interrupt continuous clinching and force another buyer to clinch more at the current price.
Consider two buyers with budgets $\bm{B}=(1,1)$ and bids $\bm{v}'=(5/2,3)$,
one unrestricted indivisible unit, and divisible supply $m=3/5$.
At $c=1/2$, both buyers satisfy $b_i=c(d_i-x_i)$,
and their bounds decrease from $2$ to $1$.
Clinching starts at $c=5/8$ and proceeds symmetrically.
Solving \eqref{eq:ode} gives
$x_i(c)=4/5-5/(16c^2)$ and $b_i(c)=5/(8c)$.
At $c=5/4$, each buyer holds $3/5$ units with remaining budget $1/2=c(1-3/5)$.
Both again satisfy the bound-decrease condition.
Buyer~$1$ comes first in the fixed order, and her bound $d_1$ falls to $0$.
The indivisible unit must go to buyer~$2$ at termination.
The integer term in \eqref{eq:clinch} makes her clinch the remaining $2/5$ of a unit at once, exhausting her budget.
The final quantities $\bm{x}=(3/5,1)$ decompose into $\bm{y}=(0,1)$ and $\bm{z}=(3/5,0)$, with $\bm{p}=(1/2,1)$.

Deferring the decomposition alone does not suffice. 
If the item is treated as divisible, the clinching auction makes the same
purchases for $c<5/4$ but continues until buyer~$1$ drops out at $c=5/2$.
It ends with $\bm{x}=(3/4,17/20)$, which no allocation implements because assigning the indivisible item to either buyer would exceed her total.
\begin{remark}
For $m=0$, the auction agrees with that of \citet{HiraiSato2025}.
For $f=0$, it reduces to the continuous form of the adaptive clinching auction~\citep{BCMX2010}. See \Cref{app:pure-equivalence}
for further details.
\end{remark}

\subsection{Feasibility and Basic Properties}\label[section]{sec:auction-feasibility}
To analyze feasibility, let $\Psi_i$ denote buyer $i$'s \emph{wishful allocation}~\citep{GMP2020}, an upper bound on her final quantity.
At $c=0$, set $\Psi_i=+\infty$ for all buyers.
For $c>0$, set $\Psi_i=x_i+b_i/c$ if $i\in A$ and $\Psi_i=x_i$ otherwise. 
Consider the total allocations satisfying these upper bounds:
\begin{align}\label{eq:completions}
\cC(\bm{d},\bm{\Psi})=\{\bm{y}+\bm{z}\mid {}&\bm{y}\in\cBZ(f),\ \bm{y}\le\bm{d},\ \bm{z}\ge0,\ \bm{z}(N)=m,\ \bm{y}+\bm{z}\le\bm{\Psi}\},
\end{align}
where $\cBZ(f)\coloneqq\{\bm{y}\in P(f)\cap\Z_+^N\mid\bm{y}(N)=l\}$
is the set of integer bases.
The following lemma gives a min--max characterization of each buyer's minimum quantity.
\begin{lemma}\label[lemma]{lem:minimum}
For integer $\bm{d}\ge0$ and $\bm{\Psi}\ge\bm{d}$, if $f_{\bm{d}}(N)=l$ and $\bm{\Psi}(N)\ge s$, then $\cC(\bm{d},\bm{\Psi})$ is nonempty and
$\min\{w_i\mid\bm{w}\in\cC(\bm{d},\bm{\Psi})\}
=\max\{\mu_i(\bm{d}),s-\bm{\Psi}(N\setminus i)\}$.
\end{lemma}
At a fixed price, clinching preserves $\bm{d}$ and $\bm{\Psi}$ and raises
each active buyer's quantity to its minimum over the fixed set
$\cC(\bm{d},\bm{\Psi})$.
The auction maintains the following invariant:
\begin{lemma}\label[lemma]{lem:invariant}
At every state with $c>0$, the bounds $\bm{d}$ and $\bm{\Psi}$ satisfy the conditions of \Cref{lem:minimum}, and every $\bm{w}\in\cC(\bm{d},\bm{\Psi})$ satisfies $\bm{w}\ge\bm{x}$.
At a stable state, 
\begin{align}\label{eq:stable}
x_i=\max\{\mu_i(\bm{d}),s-\bm{\Psi}(N\setminus i)\}\quad(i\in N).
\end{align}
\end{lemma}
At termination, \Cref{lem:invariant} and $\bm{x}(N)=s$ give $\cC(\bm{d},\bm{\Psi})=\{\bm{x}\}$.
Thus, the greedy rule selects an integer base $\bm{y}\le\min\{\bm{d},\lfloor\bm{x}\rfloor\}$ in the fixed buyer order, yielding a feasible allocation $(\bm{y},\bm{x}-\bm{y})$.
In a phase starting at $c_0>0$, finding the price $c$ at which an integer bound decreases may require solving a polynomial equation in $c/c_0$.
We assume a root oracle that, given a real polynomial of degree at most $n$ and an interval containing a unique root, returns that root exactly. We also count evaluations of $\exp$ and $\log$ as unit-cost operations.
\begin{theorem}\label[theorem]{thm:basic}
The auction is budget feasible, satisfies IC and IR, and terminates after $O(n(l+1))$ events and phases.
It uses $\operatorname{poly}(n,l)$ value queries to $f$ and real-arithmetic operations, with at most one root oracle call per~phase.
\end{theorem}
For IC, with other bids fixed, quantities and payments coincide
until the price reaches the lower of the two bids.
Underreporting can only forgo units priced at or below the true valuation,
while overreporting can only add units priced at or above it.
Neither deviation increases utility, and the final decomposition
preserves quantities and payments.
The proofs and details appear in \Cref{app:auction,app:basic-proof}.
\section{Efficiency}
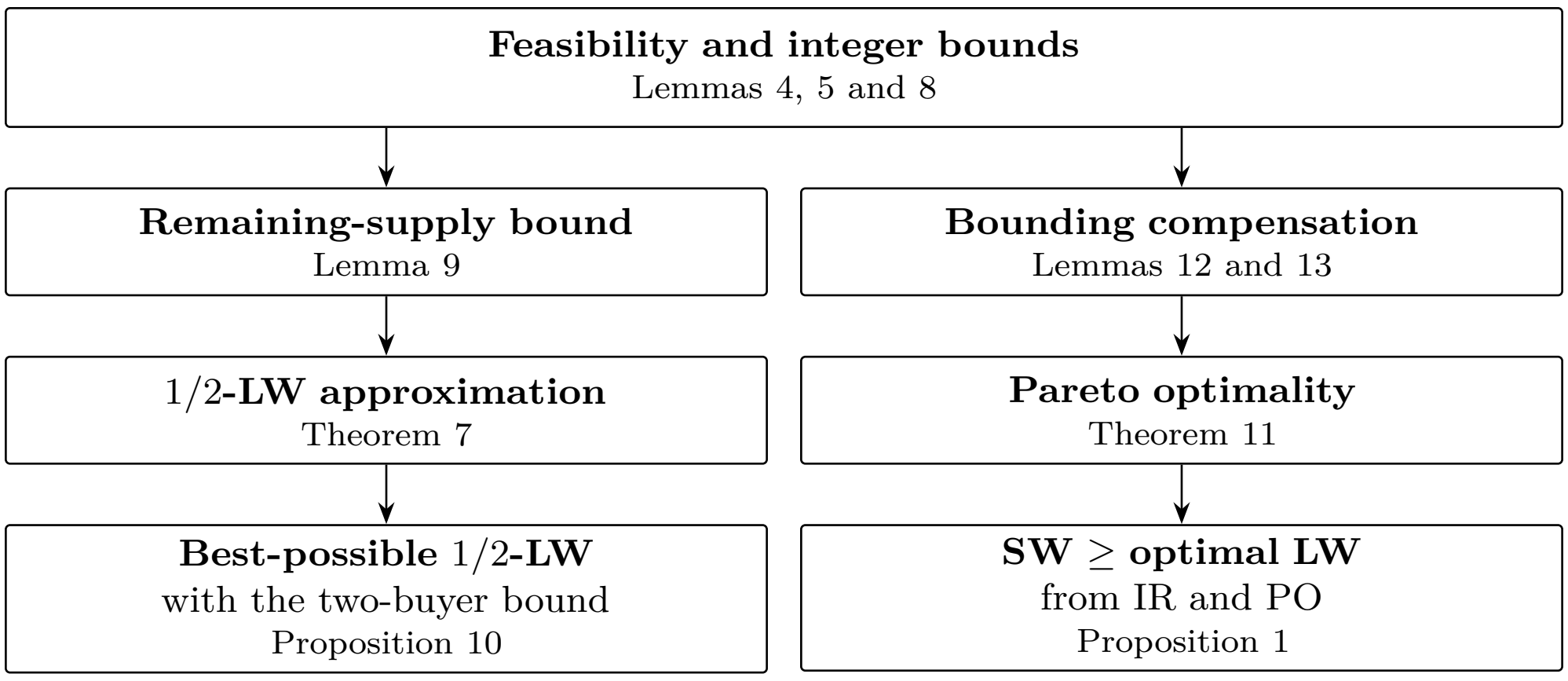
\begin{figure}[t]
\centering
\begin{tikzpicture}[
  >=Stealth, line width=.5pt,
  box/.style={draw, rounded corners=1pt, align=center,
    font=\footnotesize, inner sep=3pt, text width=6.2cm},
  root/.style={box, text width=12.9cm, minimum height=10mm},
  method/.style={box, minimum height=9mm},
  result/.style={box, minimum height=9mm},
  consequence/.style={box, minimum height=12mm}
]
\node[root] (inv)
  {\textbf{Feasibility and integer bounds}\\[-1pt]
   {\scriptsize\Cref{lem:minimum,lem:invariant,lem:cap-band}}};

\node[method,below=5mm of inv,xshift=-3.35cm] (lw)
  {\textbf{Remaining-supply bound}\\[-1pt]
   {\scriptsize\Cref{lem:remaining}}};
\node[method,below=5mm of inv,xshift=3.35cm] (po)
  {\textbf{Bounding compensation}\\[-1pt]
   {\scriptsize\Cref{lem:support-price,lem:compensation}}};

\node[result,below=5mm of lw] (lwf)
  {\textbf{$1/2$-LW approximation}\\[-1pt]
   {\scriptsize\Cref{thm:lw}}};
\node[result,below=5mm of po] (pof)
  {\textbf{Pareto optimality}\\[-1pt]
   {\scriptsize\Cref{thm:po-mixed}}};

\node[consequence,below=5mm of lwf] (opt)
  {\textbf{Best-possible $1/2$-LW}\\
   with the two-buyer bound\\[-1pt]
   {\scriptsize\Cref{thm:lower-bound}}};
\node[consequence,below=5mm of pof] (sw)
  {\textbf{SW $\ge$ optimal LW}\\
   from IR and PO\\[-1pt]
   {\scriptsize\Cref{prop:po-sw}}};

\draw[->] (inv.south -| lw.north) -- (lw.north);
\draw[->] (inv.south -| po.north) -- (po.north);
\draw[->] (lw)  -- (lwf);
\draw[->] (po)  -- (pof);
\draw[->] (lwf) -- (opt);
\draw[->] (pof) -- (sw);
\end{tikzpicture}
\caption{Roadmap of the efficiency proofs.}
\label[figure]{fig:dependencies}
\end{figure}

\label[section]{sec:efficiency}
We first prove the LW guarantee using the wishful allocations, then establish PO. \Cref{fig:dependencies} shows how these results build on each other.
Write $(\yf,\zf,\pf)$ for the final outcome, with
$\xf=\yf+\zf$ and $b_i^{\mathrm f}=B_i-p_i^{\mathrm f}$. 
Throughout this section, buyers bid truthfully, i.e., $v'_i=v_i$ for each $i\in N$.
The omitted proofs appear in \Cref{app:lw-cap-band,app:mixed-efficiency}.

\subsection{Liquid Welfare}\label[section]{sec:liquid-welfare}
\begin{theorem}\label[theorem]{thm:lw}
For every feasible allocation with total quantities $\bm{x}^*$,
the following holds:
\begin{align}\label{eq:gain-revenue}
\sum_i\bigl[\min\{v_ix_i^*,B_i\}-\min\{v_ix_i^{\mathrm f},B_i\}\bigr]_+
\le\Rev.
\end{align}
In particular, $\LWO\le\LW(\xf)+\Rev\le2\LW(\xf)$.
\end{theorem}
Our proof is much shorter than that of \citet{HiraiSato2025},
while also covering mixed goods.
Their proof uses virtual buyers and backward induction over auction events.
We derive the revenue bound without either technique, using the following
bound relating $\Psi_i$ to $d_i$.
\begin{lemma}\label[lemma]{lem:cap-band}
Let $i$ be an active buyer.
If $d_i$ has not been decreased, then $d_i=f(i)+1$.
Otherwise, 
$d_i\le\Psi_i\le d_i+1$.
In the latter case, $\Psi_i<d_i+1$ if $b_i>0$ and the price
has risen since the last decrease of $d_i$.
\end{lemma}

Fix a feasible comparison $(\bm y^*,\bm z^*)$ and extend
$\bm y^*$ so that $\bm y^*\in\cBZ(f)$, keeping $\bm z^*$ unchanged.
This cannot decrease any LW contribution.
Write $\bm x^*=\bm y^*+\bm z^*$ and define
\begin{align*}
\Delta_i
=\bigl[\min\{v_i x_i^*,B_i\}
-\min\{v_i x_i^{\mathrm f},B_i\}\bigr]_+.
\end{align*}
For $\Delta_i>0$, we have $x_i^*>x_i^{\mathrm f}$ and define
$\widehat v_i=\Delta_i/(x_i^*-x_i^{\mathrm f})$ and 
$h_i(c)=(x_i^*-x_i^{\mathrm f})\ind_{\{c<\widehat v_i\}}$. 
Otherwise, set $h_i=0$.
Then, $\int_0^\infty h_i(c)\,dc=\Delta_i$, with $\widehat v_i\le v_i$ whenever $\Delta_i>0$.
For $\Delta_i>0$, increasing buyer~$i$'s quantity from $x_i^{\mathrm f}$ to $x_i^*$ would raise her LW contribution by $v_i(x_i^*-x_i^{\mathrm f})$ without the budget cap.
With the cap, the same quantity increase adds $\Delta_i$ to her contribution.
We choose $\widehat v_i$ to be the average gain per additional unit, giving $\widehat v_i(x_i^*-x_i^{\mathrm f})=\Delta_i$.

At price $c$, write $x_i(c)$ for buyer $i$'s current total quantity
and recall that
$r(c)=s-\sum_i x_i(c)$ is the remaining supply.
Let $H(c)=\{i\mid h_i(c)>0\}$.
We omit the argument $c$ from current state variables when it is clear from the context.
The next lemma bounds the quantity needed on~$H$.
\begin{lemma}\label[lemma]{lem:remaining}
At every state of the auction,
\begin{align}\label{eq:remaining}
r(c)
\ge\sum_{i\in H(c)}\bigl(x_i^*-x_i(c)\bigr)
\ge\sum_i h_i(c).
\end{align}
\end{lemma}
\begin{proof}
The second inequality follows from $x_i(c)\le x_i^{\mathrm f}$.
The first is immediate when $H=\varnothing$ or $c=0$,
using $\bm x(0)=\bm0$ and feasibility in the latter case.
Assume $c>0$ and $H\ne\varnothing$.

We first show that each buyer $i\in H$ is active and can afford $x_i^*-x_i$ additional units at price $c$.
Fix $i\in H$. Then, $h_i(c)>0$ implies $c<\widehat v_i\leq v_i$.
Since $\Delta_i>0$, we have $v_i x_i^{\mathrm f}<B_i$.
By IR, $p_i\le p_i^{\mathrm f}\le v_i x_i^{\mathrm f}<B_i$.
Thus, $i\in A$.

The definition of $\widehat v_i$ gives
$\Delta_i>c(x_i^*-x_i^{\mathrm f})$.
Then, by $p_i\le cx_i$ and
$\Delta_i\le B_i-v_ix_i^{\mathrm f}$, we have 
\begin{align}\label{eq:budget-bound}
b_i
\ge B_i-cx_i
\ge\Delta_i+c(x_i^{\mathrm f}-x_i)
>c(x_i^*-x_i).
\end{align}
Hence, we have $\Psi_i=x_i+b_i/c>x_i^*\ge y_i^*$.

Before $d_i$ first decreases, $y_i^*\le f(i)<d_i$.
Afterwards, \Cref{lem:cap-band} and integrality give $y_i^*\le d_i$.
Setting the coordinates of $\bm y^*$ outside $H$ to zero yields a vector in $P(f_{\bm d})\cap\Z_+^N$.
Since $f_{\bm d}(N)=l$ by Lemma \ref{lem:invariant}, 
it extends to an integer base
$\bar{\bm y}\le\bm d$ with $\bar y_i\ge y_i^*$ for $i\in H$.

We now perform the case-by-case analysis:

\emph{Case 1: $x_k\le\mu_k(\bm d)$ for every $k\notin H$.}
Since $\bar y_k\ge\mu_k(\bm d)\ge x_k$ outside $H$
and $\bar y_i\ge y_i^*$ on $H$, we have 
$\bm x^*(H)+\bm x(N\setminus H)
\le \bar{\bm y}(N)+\bm z^*(H)\le l+m=s$.
Subtracting $\bm x(N)$ proves the bound.

\emph{Case 2: $x_k>\mu_k(\bm d)$ for some $k\notin H$.}
For such a buyer $k$, \Cref{lem:minimum,lem:invariant} imply
$x_k\le s-\bm\Psi(N\setminus k)$.
Therefore,
$r \ge\bm\Psi(N\setminus k)-\bm x(N\setminus k)
=\bm b(A\setminus k)/c
\ge\bm b(H)/c
>\sum_{i\in H}(x_i^*-x_i)$.
The last two inequalities use $H\subseteq A\setminus k$
and \eqref{eq:budget-bound}.
\end{proof}

\begin{proof}[Proof of \Cref{thm:lw}]
By \Cref{lem:remaining}, all $h_i$ are zero at termination,
and they remain zero afterwards because they are nonincreasing.
We also set $r(c)=0$ after termination.

Each unit contributes to $r(c)$ at every price below its purchase price.
Integrating $r(c)$ therefore gives total revenue, including payments
for purchases at discrete events:
\begin{align}\label{eq:rev-integral}
\Rev=\int_0^\infty r(c)\,dc
\ge\sum_i\int_0^\infty h_i(c)\,dc
=\sum_i\Delta_i.
\end{align}
Choose $(\bm y^*,\bm z^*)$ so that it maximizes LW over feasible allocations.
Then, $\LWO\le\LW(\xf)+\Rev\le2\LW(\xf)$,
where the last inequality uses IR and budget feasibility.
\end{proof}

This LW guarantee is best possible even without IR or PO.
\begin{proposition}\label[proposition]{thm:lower-bound}
No deterministic, budget-feasible IC mechanism guarantees more than
half of optimal LW over mixed supplies, even with two buyers
and one indivisible unit.
\end{proposition}
\begin{proof}
Fix the other bids. By the monotonicity argument in Myerson's lemma~\cite[Lemma~2]{Myerson1981}, IC implies that each buyer's total quantity is nondecreasing in her own bid.
Take budgets $(1,1)$, one indivisible unit, and divisible supply $\eps\in(0,1)$.
At valuations $(v,v)$ with $v\ge1/\eps$, some buyer $j$ receives
no indivisible unit and hence at most $\eps$ units.
Lower only her valuation to $1$.
By monotonicity, her quantity remains at most $\eps$.
The buyers' LW contributions are then at most $\eps$ and $1$,
giving LW at most $1+\eps$.
At this profile, assigning the indivisible unit to $j$ and all
divisible supply to the other buyer gives $\LWO=2$.
The ratio is therefore at most $(1+\eps)/2$,
which tends to $1/2$ as $\eps\to0$.
\end{proof}

\subsection{Pareto Optimality}\label[section]{sec:pareto-optimality}
\begin{theorem}\label[theorem]{thm:po-mixed}
The auction in \Cref{alg:main} satisfies PO.
\end{theorem}
Since the auction satisfies both IR and PO, \Cref{prop:po-sw} implies that the SW of its allocation is at least the maximum LW over all feasible allocations.

A Pareto improvement compensates buyers who lose quantity
without reducing the seller's revenue.
For an alternative outcome $(x',p')$, write
$\Delta x_i=x_i'-x_i^{\mathrm f}$ and
$\Delta p_i=p_i'-p_i^{\mathrm f}$.
A weak Pareto improvement requires
\begin{align}\label{eq:payment-change}
\Delta p_i\le\min\{b_i^{\mathrm f},v_i\Delta x_i\},
\qquad \Delta\bm p(N)\ge0.
\end{align}
Buyers who lose quantity require compensation.
We compare it with the additional payments permitted by other buyers'
remaining budgets and the value of their gains.

Suppose $m>0$. Let $\rho$ be the highest valuation among buyers
with unspent final budget, or zero if budgets are
exhausted. If buyer~$j$ receives divisible goods, every other buyer
with unspent final budget drops out at her valuation while $j$ is still
active (\Cref{lem:final-divisible}). Hence, $v_j\ge\rho$. Any buyer
who values the divisible good more than $j$ has exhausted her budget.

Consider a feasible transfer $\yf-\unit_i+\unit_j\in\cBZ(f)$ with
$v_i<\min\{v_j,\rho\}$. Since $v_i<\rho$, buyer~$i$ holds no
divisible goods. At price $v_i$, buyer~$j$ cannot afford to raise
her total quantity to $y_j^{\mathrm f}+1$ (\Cref{lem:local}).
Later purchases cost at least $v_i$ per unit and cannot close this
shortfall. Thus, $b_j^{\mathrm f}<v_i(1-z_j^{\mathrm f})$.
Even if $j$ gives all her divisible goods to $i$ for the
indivisible unit, the refund $i$ needs exceeds $j$'s remaining budget.
A third buyer, however, might pay more for those divisible goods than
$i$ values them and thereby help fund the refund. We must 
compare the refunds needed by buyers who lose quantity with the
additional payments available from all buyers who~gain.

Write $\bm x'=\bm y'+\bm z'$ for the alternative total allocation.
Then $\sum_i\min\{b_i^{\mathrm f},v_i\Delta x_i\}$ is the largest
possible revenue change that leaves every buyer at least as well off.

Fix the new integer allocation. We use a common price $\lambda$ to
value changes in divisible quantities. Each unit transferred counts
as $-\lambda$ for its former owner and $+\lambda$ for its recipient.
These amounts cancel when all $m$ units are reallocated. The next
lemma shows that the price can be chosen so that their sum bounds
the net change in revenue for every division of the divisible good.
Assume $\rho>0$ in both lemmas below.

\begin{lemma}\label[lemma]{lem:support-price}
For every $\bm y'\in\cBZ(f)$, there exists $0<\lambda\le\rho$
such that, for every $\bm z'\ge0$,
\[
\sum_i\min\{b_i^{\mathrm f},v_i\Delta x_i\}
\le \lambda\bigl(\bm z'(N)-m\bigr).
\]
If equality holds, then $\lambda=\rho$ and
$\sum_i\min\{v_i,\rho\}(y_i'-y_i^{\mathrm f})=0$.
\end{lemma}

The proof matches each gained integer unit to a lost unit through
a feasible exchange. When $\lambda<\rho$, it identifies a matched
recipient who clinches additional quantity as other buyers drop out
with unspent budgets at prices above $\lambda$.
The excess of her payments for these quantities over their cost at
$\lambda$ bounds what those buyers could contribute beyond
$\lambda$ times their quantity gains (\Cref{lem:exit-accounting}).
\Cref{app:mixed-efficiency} gives the choice of $\lambda$ and the
full proof.

When $\bm z'(N)=m$, \Cref{lem:support-price} rules out an increase
in revenue. It remains to consider the case in which the additional
payments exactly cover the refunds. Revenue then stays unchanged,
and a strict gain for one buyer without a loss for another would
require an increase in total value.

\begin{lemma}\label[lemma]{lem:compensation}
For every feasible reallocation $(\bm y',\bm z')$ with
$\bm y'(N)=l$ and $\bm z'(N)=m$, it holds 
$\sum_i\min\{b_i^{\mathrm f},v_i\Delta x_i\}\le0$.
If equality holds, then $\sum_i v_i\Delta x_i=0$.
\end{lemma}

\begin{proof}
The inequality follows from \Cref{lem:support-price} and
$\bm z'(N)=m$. Suppose equality holds. By the same lemma,
the integer reallocation preserves total value when each valuation
is capped at $\rho$. At these capped valuations, the original
divisible allocation already attains the maximum value $\rho m$,
since every recipient has valuation at least $\rho$. The total
change in capped value is therefore nonpositive, giving the last
inequality in
$0=\sum_i\min\{b_i^{\mathrm f},v_i\Delta x_i\}
\le \sum_i\min\{v_i,\rho\}\Delta x_i \le 0$.
The first inequality holds for each buyer separately. For
$v_i\le\rho$, the payment bound is at most $v_i\Delta x_i$.
For $v_i>\rho$, the budget is exhausted and
$\min\{0,v_i\Delta x_i\}\le\rho\Delta x_i$, strictly unless
$\Delta x_i=0$. Equality throughout therefore forces every buyer
with $v_i>\rho$ to keep her original quantity. Removing the cap
leaves the total value change unchanged, giving
$\sum_i v_i\Delta x_i=0$.
\end{proof}

\begin{proof}[Proof of \Cref{thm:po-mixed}]
When $m=0$, our auction yields the same outcome as the
clinching auction of \citet{HiraiSato2025}
(\Cref{app:pure-equivalence}). Their guarantee thus applies.
Assume $m>0$ and consider a weak Pareto improvement.
Extend its allocation to use all supply without changing payments.

If $\rho=0$, every buyer has exhausted her budget.
Budget feasibility gives $\Delta p_i\le0$ for all $i$.
Together with \eqref{eq:payment-change}, this forces every payment to remain unchanged.
With payments unchanged, no buyer can receive less without losing utility.
Since both allocations use all supply, every buyer receives her original quantity.
If $\rho>0$, \Cref{lem:compensation} and \eqref{eq:payment-change} give
$0\le\Delta\bm p(N)
\le\sum_i\min\{b_i^{\mathrm f},v_i\Delta x_i\}\le0$.
Equality in \Cref{lem:compensation} also gives $\sum_i v_i\Delta x_i=0$.
Thus, both revenue and SW are unchanged.
Since SW equals total buyer utility plus seller revenue,
no participant can strictly improve while all participants weakly improve.
\end{proof}

\section{Restricted Access to Divisible Goods}
\label[section]{sec:alternative-constraints}

Our main theorem assumes that the divisible supply is available to
every buyer. This assumption cannot be dropped in general: restricting
a single buyer's access to the divisible good can make IC, IR, and PO
incompatible, even when both types of goods are subject to matroid
constraints. Consider three buyers with public budgets
$\bm B=(2,3/2,2)$, one unrestricted indivisible item, and one
divisible unit available only to buyers~1 and~2.
\begin{theorem}\label[theorem]{thm:eligibility-impossible}
No deterministic mechanism for this market with budget-feasible,
nonnegative payments satisfies IC, IR, and PO for all positive
valuation profiles.
\end{theorem}

\emph{Proof sketch.}
Fix $(v_2,v_3)=(17/16,1)$. IC across buyer~2's reports restricts her
payments. Together with PO, these restrictions force buyer~2 to receive
the item and buyer~1 a positive divisible quantity when buyer~1 reports
$11/20$. When buyer~1 raises her report to $3/4$, buyer~3 must receive
the item, and buyer~2 must receive the entire divisible unit.
Buyer~1's quantity falls to zero, contradicting IC.
\Cref{app:eligibility-impossible} gives the full proof and an
impossibility instance for every cap $\eta\in[0,1)$ on buyer~3's
share of the divisible good.

\Cref{app:constraint-transfer} also gives sufficient conditions for
implementing divisible clinching outcomes under a common polymatroid
constraint on total quantities or separate constraints on the two
types of goods, while preserving the guarantees of our main theorem.

\section{Concluding Remarks}\label[section]{sec:conclusion}
Our mixed-goods clinching auction determines total quantities and payments
as the price rises, deferring the decomposition into divisible and indivisible
goods until termination. With public budgets and common divisible supply,
it satisfies IC, IR, and PO. Its $1/2$-approximation to optimal LW is best
possible among deterministic, budget-feasible IC mechanisms. We also give
sufficient conditions for implementing divisible auction outcomes with mixed
goods. For any $\eta\in(0,1)$, capping one buyer's share of the divisible
supply at $\eta$ can make IC, IR, and PO incompatible for deterministic
mechanisms with budget-feasible, nonnegative payments, even with three buyers.
It remains open which guarantees extend to buyer-specific or privately known
value ratios and how closely optimal LW can be approximated without PO when
access to divisible goods is restricted.

\label[section]{main:end}

\section*{Acknowledgements}
    Ryosuke Sato was supported by JST ERATO Grant Number JPMJER2301 and JSPS KAKENHI Grant Number JP26K21174.

We used OpenAI's GPT and Anthropic's Claude to assist with developing
research ideas, constructing and checking proofs, and drafting and
revising the manuscript.
The author takes full responsibility for the content of this paper,
including all mathematical claims and proofs.

\clearpage

\bibliography{references}

\ifwithappendices
\clearpage

\appendix

\section{Reduction by Initial Allocation}
\label[appendix]{app:mandatory-quantities}
For general $f$, set $a_i\coloneqq l-f(N\setminus i)$,
$\widetilde f(S)\coloneqq f(S)-\bm a(S)$, and $\widetilde l=\widetilde f(N)$.
By submodularity, $a_i$ is the smallest marginal contribution of $i$.
Thus $\widetilde f$ is normalized, monotone, integer-valued, and
submodular, with $\widetilde f(N\setminus i)=\widetilde l$ for every $i$.
Every integer base satisfies $y_i=l-\bm y(N\setminus i)\ge a_i$.
Translating the base constraints therefore gives 
$\cBZ(f)=\bm a+\cBZ(\widetilde f)$.

Allocate $\bm a$ for free and run the auction with $\widetilde f$,
divisible supply $m$, and the original bids and budgets.
If $\widetilde l+m=0$, take the residual allocation and payments to be zero.
Since $\widetilde l\le l$, the same running-time bound applies.
Adding $\bm a$ to the residual indivisible allocation preserves feasibility
and leaves payments unchanged. Each buyer's utility increases by $v_i a_i$
regardless of her report, preserving IC and IR. 

To see that PO also carries
over, extend any proposed Pareto improvement to a full allocation at the
same payments. Subtracting $\bm a$ from the indivisible part of any full
original allocation gives a feasible residual allocation. Applying this
to both the auction outcome and the proposed improvement preserves utility
differences and revenue, contradicting PO in the residual instance.

For LW, extend an arbitrary feasible comparison allocation to full supply without decreasing any buyer's quantity.
Let $\bm u$ and $\bm x$ be the residual total quantities of
the extended allocation and the auction outcome, respectively.
Since $\varphi_i(q)=\min\{v_iq,B_i\}$ is nondecreasing and concave,
\[
[\varphi_i(a_i+u_i)-\varphi_i(a_i+x_i)]_+
\le [\varphi_i(u_i)-\varphi_i(x_i)]_+.
\]
Summing over buyers and applying \eqref{eq:gain-revenue} to the residual
auction gives the same revenue bound, and hence the same
$1/2$-approximation, for the original instance.

\section{Proof of \Cref{prop:fluiddecrease}}
\label[appendix]{app:polimits}

Consider three buyers, each with budget $B>0$.
The supply consists of one unrestricted indivisible item and $m\in(0,1)$ units of a divisible good. 
Fix $\varepsilon\in(0,m)$ and a deterministic, budget-feasible mechanism with nonnegative payments satisfying IC, IR, and PO. We use profiles with distinct valuations in $(0,B)$ to exhibit a decrease of more than $m-\varepsilon$ in one buyer's divisible allocation.
Since $v_k>0$ for every $k\in N$, PO requires full allocation of both supplies.

Let $i$ be the highest-valued buyer.
Suppose some buyer $k\in N\setminus i$ holds the indivisible item.
Give the item to $i$ and all her $z_i$ divisible units to $k$, and transfer $v_k(1-z_i)$ from $i$ to $k$.
By IR, buyer~$i$'s new payment is at most $v_i z_i+v_k(1-z_i)<v_i<B$, since $z_i\le m<1$.
The exchange strictly benefits buyer~$i$ while preserving buyer~$k$'s utility and the revenue, contradicting PO.
Thus, buyer~$i$ must receive the item.

The middle-valued buyer $j$ holds no indivisible item and has unspent budget, since IR gives $p_j\le v_jm<B$.
If the lowest-valued buyer held any divisible goods, $j$ could buy a small quantity at that buyer's valuation and strictly gain. The lowest-valued buyer receives nothing and pays zero by IR.

Choose $\underline v\in(B/(1+\varepsilon/2),B)$ and consider distinct
bids in $(\underline v,B)$. Suppose the winner of the item receives $z>0$
divisible units and pays $p$. At report $\underline v$, she is the lowest bidder and therefore
receives and pays zero. If her true valuation were $\underline v$,
IC would give $0\ge\underline v(1+z)-p$. Since $p\le B$, we obtain
$z\le B/\underline v-1<\varepsilon/2$. 
Now raise only the middle buyer's bid above the highest bid, keeping it
below $B$. Initially she receives more than $m-\varepsilon/2$ divisible
units because the winner receives less than $\varepsilon/2$ and the
lowest-valued buyer receives nothing. After raising her bid, she wins
the item and receives less than $\varepsilon/2$ divisible units.
Her divisible allocation thus falls by more than $m-\varepsilon$.

\section{Equivalence in the Pure Cases}\label[appendix]{app:pure-equivalence}
This section shows that our auction recovers the polyhedral clinching auction
for indivisible goods when $m=0$ and the adaptive clinching auction when $f=0$.

\subsection{Indivisible goods ($m=0$).}
Let $\hat d_i\coloneqq d_i-x_i$ denote buyer~$i$'s remaining demand.
With this notation, the initial demands, budget events $b_i/c=\hat d_i>0$,
and exits at reported valuations agree with those of \citet{HiraiSato2025}.
For general $f$, first apply the reduction in \Cref{app:mandatory-quantities}.

Since $m=0$ and $\bm d\le\bm\Psi$, the feasible totals are the integer
bases $\bm y\le\bm d$. By \Cref{lem:minimum,lem:invariant}, a fixed-price
clinch raises each active buyer's quantity to its minimum over these
bases, $\mu_i(\bm d)$. At a stable state, each such buyer has $x_i=\mu_i(\bm d)$.
While $\bm d$ is fixed, the invariant still requires
$x_i\le\mu_i(\bm d)$, ruling out further purchases as the price rises.

At any common state before clinching, substituting $\hat{\bm{d}}=\bm{d}-\bm{x}$
into Hirai and Sato's formula for the rank function~gives
\[
f_{\bm{x},\hat{\bm{d}}}(S)
=\min_{T\subseteq S}\{f(T)-\bm{x}(T)+\hat{\bm{d}}(S\setminus T)\}
=f_{\bm{d}}(S)-\bm{x}(S).
\]
Since $f_{\bm{d}}(N)=l$, their clinching amount is
\[
f_{\bm{x},\hat{\bm{d}}}(N)-f_{\bm{x},\hat{\bm{d}}}(N\setminus i)
=\mu_i(\bm{d})-x_i,
\]
which is exactly the quantity buyer~$i$ clinches.
A buyer with $\hat d_i=0$ cannot clinch, and $d_i=x_i$ still holds when she drops out.
With the same tie-breaking order, each buyer clinches the same quantity at each price in both auctions.
Therefore, their PO guarantee yields \Cref{thm:po-mixed} for $m=0$.

\subsection{Divisible goods ($f=0$).}
Here, $\bm y=\bm 0$ is the only indivisible allocation and
$\mu_i(\bm d)=0$ for every $i$. The fixed-price rule reduces to 
$\delta_i=\max\{0,r-\bm b(A\setminus i)/c\}$.
Updating an integer bound alone therefore leaves quantities and
payments unchanged. This fixed-price rule, the price dynamics
in \eqref{eq:ode}, and the value-exit rule coincide with those of the
continuous adaptive clinching auction~\citep{BCMX2010}.

\section{Mixed Implementation under Alternative Constraints}
\label[appendix]{app:constraint-transfer}

We consider two types of polymatroid constraints.
For a normalized, monotone submodular function $h$, write
$\mathcal B(h)=\{\bm x\in P(h)\mid\bm x(N)=h(N)\}$.
\begin{itemize}
\item \emph{A common constraint on total quantities.}
The total allocation satisfies $\bm x=\bm y+\bm z\in P(h)$,
where $h(N)=s$.
There are no separate constraints beyond the supplies.
\item \emph{Separate constraints on the two types of goods.}
Indivisible allocations satisfy $\bm y\in P(f)\cap\Z_+^N$ as before.
Divisible allocations satisfy $\bm z\in P(g)$, where $g$ is
normalized, monotone, and submodular with $g(N)=m$.
\end{itemize}

The divisible polyhedral clinching auction runs on $P(h)$ under
the common constraint and on $P(f+g)$ under separate constraints.
Assume it is deterministic and budget feasible, satisfies IC, IR,
and PO, and achieves a $1/2$-approximation to the relaxed LW optimum
under any required price-step assumptions~\citep{GMP2015,Sato2026}.
We seek a mixed allocation with the same total quantities and payments.

\begin{proposition}\label[proposition]{prop:constraint-transfer}
Under either condition below, the divisible polyhedral clinching
auction admits a deterministic mixed implementation that is budget
feasible, satisfies IC, IR, and PO, and achieves a $1/2$-approximation
to optimal LW.
\begin{enumerate}
\item[(i)] Under a common constraint on total quantities, every output
satisfies $\sum_i\lfloor x_i\rfloor\ge l$.
In particular, $m\ge n-1$ suffices.
\item[(ii)] Under separate constraints, every relaxed base is implementable:
$\cBZ(f)+\mathcal B(g)=\mathcal B(f+g)$.
\end{enumerate}
\end{proposition}
Here, $+$ denotes the Minkowski sum.

\begin{proof}
Since valuations are positive and the divisible auction satisfies PO,
every output $\bm x$ is a base. 
Under a common constraint, implementing $\bm x$ requires
$\bm y(N)=l$ and $y_i\le\lfloor x_i\rfloor$, hence the floor condition.
Conversely, assign $l$ integer units within these bounds
in a fixed buyer order and set $\bm z=\bm x-\bm y$.
Then, $\bm z\ge0$, $\bm z(N)=m$, and $\bm y+\bm z=\bm x$
satisfies the common constraint.
If $m\ge n-1$, then
$\sum_i\lfloor x_i\rfloor>\bm x(N)-n=l+m-n\ge l-1$.
The left side is integer and therefore at least $l$.
Under separate constraints, $\bm x\in\mathcal B(f+g)$.
Condition~(ii) gives $\bm y\in\cBZ(f)$ and $\bm z\in\mathcal B(g)$
such that $\bm x=\bm y+\bm z$.
Choose such a decomposition by a fixed deterministic rule.
This gives a feasible mixed allocation with total quantities $\bm x$.

In both cases, implementation preserves quantities and payments
at every report profile, hence IC, IR, and budget feasibility.
Any mixed Pareto improvement would also improve the divisible outcome,
which is Pareto optimal.
Since implementation preserves LW and the relaxed optimum is at least
the mixed optimum, the $1/2$ guarantee carries over.
\end{proof}

With two buyers, one unrestricted indivisible item, and common divisible
supply $m\ge1$, every relaxed allocation of the full supply is implementable.
Since the quantities sum to $m+1\ge2$, one buyer receives at least one unit.
Assign the item to that buyer and split the divisible good to match
each buyer's relaxed total.
For general $g$, monotonicity gives $g(\{i\})\le g(N)=m$ for $i=1,2$.
Buyer~1 receives at most $g(\{1\})$ units without the item and at least
$1+m-g(\{2\})$ with it. Condition~(ii) holds exactly when these ranges
leave no gap, i.e., $g(\{1\})+g(\{2\})\ge m+1$.
Thus, $m\ge1$ is necessary for condition~(ii) and sufficient under
common supply, where $g(\{1\})=g(\{2\})=m$.

\section{Proof of \Cref{lem:minimum,lem:invariant,lem:cap-band}}
\label[appendix]{app:auction}
Recall that \(s=l+m\) is the total supply. We start with $\bm x=\bm p=\bm 0$
and assume $f(N)=f(N\setminus i)=l$ for every $i$.
For general $f$, first apply the reduction in \Cref{app:mandatory-quantities}.

\subsection{Proof of Lemma~\ref{lem:minimum}}
\label[appendix]{app:completion-proofs}
Fix a buyer $i$. Since $f_{\bm d}(N)=l$, placing $i$ last in the
greedy order gives an integer base $\bm y\le\bm d$ with
$y_i=l-f_{\bm d}(N\setminus i)=\mu_i(\bm d)$.
The capacities $\bm\Psi-\bm y$ are nonnegative and sum to at least $m$.
Fill the capacities outside $i$ first and give $i$ the remaining
divisible supply. Her total quantity is
\[
y_i+[m-(\bm\Psi-\bm y)(N\setminus i)]_+
=\max\{\mu_i(\bm d),s-\bm\Psi(N\setminus i)\}.
\]
For every allocation in $\cC(\bm{d},\bm{\Psi})$, both terms are lower bounds on buyer~$i$'s total quantity.
Thus, $\cC(\bm{d},\bm{\Psi})$ is nonempty, and the right-hand side gives the minimum buyer~$i$ can receive.

\subsection{Proof of \Cref{lem:invariant}}
Under the conditions of \Cref{lem:minimum}, \eqref{eq:stable} says that $x_i$ is buyer~$i$'s minimum total quantity over $\cC(\bm{d},\bm{\Psi})$.
In particular, every allocation $\bm{w}$ in this set satisfies $\bm{w}\geq\bm{x}$.

Initially, the bounds $d_i=f(i)+1$ give $f_{\bm{d}}=f$, hence $f_{\bm{d}}(N)=l$ and $\mu_i(\bm{d})=0=x_i$.
Since bids and budgets are positive, for sufficiently small $c>0$, all buyers are active, $\Psi_i=B_i/c>d_i$, and $\bm{\Psi}(N\setminus i)>s$ for every $i$.
Thus, $\bm{\Psi}(N)>s$, satisfying the conditions of \Cref{lem:minimum}.
Since the second term in \eqref{eq:stable} is negative, the maximum equals $x_i=0$.

\emph{Updates at a fixed price.}
Suppose buyer~$i$ drops out at her valuation or her bound decreases by one.
By the induction hypothesis, \eqref{eq:stable} holds just before the update.
By \Cref{lem:minimum}, 
some feasible allocation gives her total quantity $x_i$.
Let $y_i$ be its indivisible part.
If $i$ drops out, this allocation respects the new bounds because $y_i\le\lfloor x_i\rfloor$ and her new wishful allocation is $x_i$.
Just before the bound decreases,  $y_i\le x_i<\Psi_i=d_i$, giving $y_i\le d_i-1$.
Thus, the updated set is a nonempty subset of the old one, and each of its allocations gives every buyer~$j$ at least $x_j$.
In particular, $l=\bm{y}(N)\le f_{\bm{d}}(N)\le l$ and $\bm{\Psi}(N)\ge s$.
Both updates preserve $\bm{d}\le\bm{\Psi}$.

Purchases leave $\bm{d}$ and $\bm{\Psi}$ unchanged.
By \Cref{lem:minimum} and \eqref{eq:clinch}, each active buyer clinches exactly enough to reach her new minimum.
This increment is at most $\Psi_i-x_i=b_i/c$.
For each inactive buyer~$i$, every allocation in $\cC(\bm{d},\bm{\Psi})$ gives her exactly $x_i$, since $\Psi_i=x_i$.
This restores \eqref{eq:stable} for all buyers.
If buyer~$i$ exhausts her budget, removing her from the active set preserves both bounds because $\Psi_i=x_i$ and $d_i\le\lfloor x_i\rfloor$.

\emph{As the price rises.}
The budget term in \eqref{eq:stable} can be written as
\begin{align*}
s-\bm{\Psi}(N\setminus i)=x_i+r-\bm{b}(A\setminus i)/c.
\end{align*}
For $i\in C$, \eqref{eq:ode} preserves $\bm{b}(A\setminus i)=cr$, since both sides have derivative $(1-|C|)r$.
For active buyers outside $C$, the other active budgets exceed $cr$ until the next entry into $C$.
For inactive $i$, $\bm{b}(A\setminus i)=\bm{b}(A)>cr$, since some $j\in A$ has $b_j>0$ and $\bm{b}(A\setminus j)\ge cr$.
Consequently, $\bm{\Psi}(N)=\bm{x}(N)+\bm{b}(A)/c\ge s$.
Throughout the phase, $\bm{d}$ is fixed, hence $f_{\bm{d}}(N)=l$.
The phase ends if an active buyer's $\Psi_i$ reaches $d_i$, ensuring $\bm{d}\le\bm{\Psi}$ throughout.

For $i\in C$, the budget term $s-\bm{\Psi}(N\setminus i)$ in \eqref{eq:stable} stays equal to $x_i$.
The integer term $\mu_i(\bm{d})$ is fixed and is at most $x_i$ at the start.
Since $x_i$ can only increase, the budget term remains the maximum.
For $i\notin C$, the budget term is strictly below $x_i$ at the start, hence initial stability gives $x_i=\mu_i(\bm{d})$.
Both $x_i$ and $\mu_i(\bm{d})$ stay fixed, and the budget term stays at most $x_i$.
Thus, the integer term remains the maximum.
Therefore, \eqref{eq:stable} holds throughout the phase and, by continuity, at its endpoint before the next~update.

At termination, every allocation in $\cC(\bm{d},\bm{\Psi})$ is componentwise at least $\bm{x}$ and has total quantity $s=\bm{x}(N)$.
The set is thus $\{\bm{x}\}$, and the greedy rule gives a feasible decomposition.

\subsection{Proof of \Cref{lem:cap-band}}
\label[appendix]{app:lw-cap-band}
Fix an active buyer $i$. At a fixed price, purchases preserve $\Psi_i=x_i+b_i/c$.
Between events, \Cref{eq:ode} gives $b_i'=-cx_i'$ and hence $\Psi_i'=-b_i/c^2$.
Thus, $\Psi_i$ is nonincreasing and decreases strictly as the price rises while $b_i>0$.
Initially, $d_i=f(i)+1$. Each unit decrease occurs when $\Psi_i$ equals the old bound, leaving $\Psi_i=d_i+1$ for the new bound.
Hence, $\Psi_i\le d_i+1$ after the first decrease, with strict inequality if $b_i>0$ and the price has risen since the last decrease.
The lower bound $d_i\le\Psi_i$ follows from \Cref{lem:invariant}.

\section{Proof of \Cref{thm:basic}}
\label[appendix]{app:basic-proof}

\subsection{Price Phases and Event Count}
\label[appendix]{app:price-phases}
\label[appendix]{app:event-count}
Between events, we express each buyer's quantity and payment as functions of the price. We use these expressions to determine the price of the next event.
Here, we show that the auction terminates after $O(n(l+1))$ events.
For a phase starting at $c_0>0$ with fixed clinching set $C$, let $k\coloneqq|C|$, $r_0\coloneqq r(c_0)$, and $t\coloneqq c/c_0$.
If $k=0$, no purchases occur.
Otherwise, integrating \eqref{eq:ode} gives $r(c)=r_0t^{-k}$ and, for each $i\in C$,
\begin{align*}
x_i(c)-x_i(c_0)&=(r_0-r(c))/k,\\
p_i(c)-p_i(c_0)&=
\begin{cases}
r_0c_0\log t,&k=1,\\
\dfrac{r_0c_0}{k-1}(1-t^{1-k}),&k\ge2.
\end{cases}
\end{align*}

For each $i\in C$, the equality $\bm{b}(A\setminus i)=cr$ persists during a phase because both sides have derivative $(1-|C|)r$ by \eqref{eq:ode}.

Suppose $i\in C$ just before another buyer drops out or the auction lowers some $d_j$ by one, and suppose $i$ remains active.
Before purchases, lowering $d_j$ preserves $\bm{b}(A\setminus i)=cr$, while another buyer dropping out can only reduce $\bm{b}(A\setminus i)$.
If $i$ buys $r-\bm{b}(A\setminus i)/c$ units at $c$, her purchase brings $cr$ down to $\bm{b}(A\setminus i)$.
Each purchase by another buyer reduces both sides by the amount paid.
Thus, $i$ can leave $C$ only if the integer term in \eqref{eq:clinch} makes her buy more.
Her new $\mu_i(\bm{d})$ exceeds her quantity before the purchase, which was at least her old $\mu_i(\bm{d})$.

Each time a buyer drops out of $C$ while remaining active, $\sum_j\mu_j(\bm{d})$ increases by at least one, since all $\mu_j(\bm{d})$ are integers and none decreases as the bounds tighten.
Any integer base within the bounds gives $\sum_j\mu_j(\bm{d})\le l$, hence this can happen at most $l$ times.
There are at most $n$ first entries into $C$, and each reentry requires a buyer to have dropped out of $C$ while remaining active.
Thus, there are at most $n+l$ entries.

The auction lowers an integer bound by one at most $\sum_j(f(j)+1)\le n(l+1)$ times and removes at most $n$ buyers.
The price rises until a buyer leaves the active set, a buyer enters $C$, or the auction lowers an integer bound.
Thus, there are $O(n(l+1))$ events and intervals of rising prices.
Once all buyers have left the active set, \Cref{lem:invariant} gives $s\le\bm{\Psi}(N)=\bm{x}(N)\le s$.
The auction therefore ends by the highest reported valuation.

\subsection{Implementation and Incentive Guarantees}
\label[appendix]{app:exact-implementation}
To prove \Cref{thm:basic}, we find the next event price in polynomial time and establish IC by comparing purchases under two bids.
When $C=\{i\}$, we group successive decreases of $d_i$.
When $|C|\ge2$, we check the bound of a buyer in $C$ with the smallest gap $\Psi_i-d_i$.

We charge unit cost for oracle queries to $f$, arithmetic operations, comparisons, $\exp$, and $\log$.
Given a polynomial of degree $k$ with real coefficients and an interval containing exactly one of its roots, we assume that this root can be found exactly in time polynomial in $k$.

\subsubsection{Bound decreases if $|C|=1$.}
Suppose $C=\{i\}$ and $\Psi_i=d_i$ with $b_i>0$ in a nonterminal stable state.
Write $\mu_j\coloneqq\mu_j(\bm{d})$ for buyer~$j$'s minimum before the decrease.
A base with $y_i=\mu_i$ satisfies the new bound because $\mu_i\le x_i<d_i$.

Fix $j\ne i$. Stability gives $x_j=\mu_j$.
If $j$ is active, then $b_j<b_i$ because $j\notin C$. If $j$ is inactive, then $\Psi_j=x_j$.
In either case,
$d_j-\mu_j\le\Psi_j-x_j<b_i/c=d_i-x_i\le d_i-\mu_i$.
The greedy base placing $i$ second to last and $j$ last minimizes both $y_i+y_j$ and $y_j$.
Comparing its sum $y_i+y_j$ with that of a base attaining $y_i=\mu_i$ gives
$y_i+\mu_j\le\mu_i+d_j<d_i+\mu_j$. 
Thus, $y_i<d_i$, and this base still attains $\mu_j$ after the decrease.
Every buyer's minimum is therefore unchanged, and the update triggers no purchase or change in $C$.

Postpone these decreases until the next valuation drop, another buyer's bound event, or an entry into $C$.
Just before processing events at that price $h$, set $d_i\gets\min\{d_i,\lfloor\Psi_i(h^-)\rfloor\}$.
This counts crossings strictly before $h$ and leaves any equality at $h$ to the fixed event order.

\subsubsection{Locating the next event.}
Buyers drop out at their reported valuations.
An active buyer $j\notin C$ reaches her bound at $b_j/(d_j-x_j)$ if $d_j>x_j$.
When $C=\varnothing$, the first entry price is $(\bm{b}(A)-\max_{j\in A}b_j)/r$.
These formulas also apply at $c=0$.
When $C$ and $A\setminus C$ are both nonempty, the next entry occurs when the common remaining budget in $C$ falls to $\max_{j\in A\setminus C}b_j$.
Inverting the payment formula in \Cref{app:price-phases} gives this price.

Let $h$ be the earliest of these prices.
If $C=\varnothing$, advance to $h$ without purchases.
If $|C|=1$, advance to $h$ using the preceding bound update.

For $k=|C|\ge2$, buyers in $C$ have equal remaining budgets and equal derivatives $\Psi_i'=-b_i/c^2$ by \Cref{app:lw-cap-band}.
Use the phase formula to evaluate the gap at $h$ for a buyer $i\in C$ minimizing $\Psi_i(c_0)-d_i$.
If the gap is nonnegative, advance to $h$ and process ties in the fixed order.
Otherwise, the next event occurs at $c_0t$, where $t\in(1,h/c_0)$ solves
\begin{align}
k(k-1)\left(d_i-x_i(c_0)-\frac{r_0}{k}\right)t^k
+\frac{k\bm{b}(A\setminus C)}{c_0}t^{k-1}=r_0.
\label{eq:cap-polynomial}
\end{align}
This follows by substituting the phase formulas and $(k-1)b_i(c)+\bm{b}(A\setminus C)=cr(c)$ into $\Psi_i(c)=d_i$. 
Up to $h$, the common budget is at least $\max_{j\in A\setminus C}b_j>0$ if $A\ne C$.
If $A=C$, $cr(c)>0$ gives $k\ge2$ and $b_i(c)=cr(c)/(k-1)>0$.
Thus, no budget is exhausted between events, and strict decrease of $\Psi_i$ gives uniqueness in \eqref{eq:cap-polynomial}.
A buyer who exhausts her budget at a fixed price is removed at that price.

\subsubsection{Running time.}
Each fixed-price update uses at most $n$ evaluations of $f_{\bm{d}}$.
To evaluate $f_{\bm{d}}(S)$, we minimize the submodular function $f(T)+\bm{d}(S\setminus T)$ over $T\subseteq S$ using polynomially many oracle queries and arithmetic operations~\citep{Schrijver2000,IwataFleischerFujishige2001}.
By \Cref{app:event-count}, the updates require $O(n^2(l+1))$ such minimizations in total.
Each phase scans the buyers and selects at most one root of a polynomial of degree at most $n$.
Positive fractional powers use $\exp$ and $\log$, and truncated floors require only comparisons with integers from $0$ to $f(i)+1$.
The final greedy allocation uses polynomially many rank evaluations.

\begin{proof}[Proof of \Cref{thm:basic}]
By \Cref{lem:invariant}, the final decomposition is feasible and preserves total quantities and payments.
Every purchase is affordable and made at a nonnegative price, giving $0\le p_i\le B_i$ for every buyer~$i$.

Fix the other bids. Under the fixed event order, buyer~$i$'s purchases at a lower bid form an initial segment of her purchases at a higher bid, even when events coincide.
Each unit she buys truthfully costs at most her valuation.
Bidding lower can only shorten this sequence, and bidding higher can only add units priced at or above her valuation.
IC and~IR~follow.

The fixed order and greedy rule uniquely determine the outcome.
By \Cref{app:event-count}, the auction terminates after $O(n(l+1))$ events and price phases.
The implementation runs in time polynomial in $n$ and $l$ under our assumption.
\end{proof}

\section{Proof of \Cref{lem:support-price}}
\label[appendix]{app:mixed-efficiency}

Throughout this section, buyers report truthfully. We write $x_i(c)$
and $b_i(c)$ for buyer~$i$'s total quantity and remaining budget after
all events at price $c$. For prices above the terminal price, we use
their final values.

\subsection{Remaining Budgets and Final Allocations}

We first consider a buyer whose current quantity exceeds her minimum
indivisible allocation under the current bounds.

\begin{lemma}\label[lemma]{lem:budget-equality}
Suppose $x_i>\mu_i(\bm d)$ at price $c>0$. If $i$ is active, then
$\bm b(A\setminus i)\le cr$, with equality at a stable state.
If $i$ is inactive, purchases at $c$ exhaust the remaining supply
and all active budgets.
\end{lemma}

\begin{proof}
By \Cref{lem:minimum,lem:invariant},
$x_i\le s-\bm\Psi(N\setminus i)=x_i+r-\bm b(A\setminus i)/c$, 
which gives $\bm b(A\setminus i)\le cr$. If $i$ is active and the
state is stable, \eqref{eq:clinch} gives
$r-\bm b(A\setminus i)/c\le0$, proving equality.

If $i$ is inactive, the upper bound becomes $\bm b(A)\le cr$.
The reverse inequality follows from \Cref{lem:invariant}, since
$\bm x(N)+\bm b(A)/c=\bm\Psi(N)\ge s$. Thus $\bm b(A)=cr$,
and \eqref{eq:clinch} gives each active buyer~$j$ exactly $b_j/c$
units. These quantities sum to $r$, and each buyer spends her
remaining budget.
\end{proof}

For the rest of this section, assume $m>0$. We now use
\Cref{lem:budget-equality} to relate the valuations of buyers who
receive divisible goods to those of buyers who retain unspent budget.

\begin{lemma}\label[lemma]{lem:final-divisible}
If $z_j^{\mathrm f}>0$, the auction ends by price $v_j$, and every
other buyer with unspent final budget drops out at her valuation
while $j$ is active.
\end{lemma}

\begin{proof}
The final base respects every earlier bound, giving
$\mu_j(\bm d)\le y_j^{\mathrm f}<x_j^{\mathrm f}$.
If $j$ drops out, she then holds $x_j=x_j^{\mathrm f}$, and
\Cref{lem:budget-equality} implies that purchases at that price
exhaust the remaining supply and all still-active budgets.
If $j$ remains active until termination, the same lemma gives
$\bm b(A\setminus j)\le cr=0$ there. In either case, the auction
ends by $v_j$, and every other buyer who retains budget must have
dropped out at her valuation while $j$ was active.
\end{proof}

\subsection{Feasible Exchanges and Dropping Out}

Assume $\rho>0$ and fix a feasible exchange
$\yf-\unit_i+\unit_j\in\cBZ(f)$ with $v_i<\min\{v_j,\rho\}$.
Then $z_i^{\mathrm f}=0$ by \Cref{lem:final-divisible}.

\begin{lemma}\label[lemma]{lem:local}
At price $v_i$, buyer~$j$ satisfies
$b_j(v_i)+v_i\bigl(x_j(v_i)-y_j^{\mathrm f}\bigr)<v_i$.
\end{lemma}

\begin{proof}
At $v_i$, consider the state after all valuation dropouts and their
purchases but before any unit decrease of a bound. If the auction
ends earlier, use the terminal state. It suffices to prove the
inequality in this state, since any subsequent purchase at $v_i$
preserves $b_j+v_i x_j$.

Buyer~$i$ now holds $x_i=x_i^{\mathrm f}=y_i^{\mathrm f}$.
Since $v_i<\rho$, some buyer~$h$ with $v_h>v_i$ is still active
and has unspent final budget. We claim that $d_j=y_j^{\mathrm f}$.
If $d_j>y_j^{\mathrm f}$, integrality implies that the exchanged
base also respects the current bounds. This gives
$\mu_i(\bm d)\le y_i^{\mathrm f}-1<x_i$. If $i$ has left,
\Cref{lem:budget-equality} would exhaust $h$'s budget. If $i$ is
still active, the auction has ended, and the same lemma gives
$\bm b(A\setminus i)\le cr=0$. Both cases contradict $h$'s
unspent budget.

If $j$ has no budget left in the chosen state, she exhausted it
while $h$ was still active. By \Cref{lem:final-divisible},
$z_j^{\mathrm f}=0$, giving $x_j=x_j^{\mathrm f}=y_j^{\mathrm f}$
as required. Otherwise, $j$ is active and her bound
$d_j=y_j^{\mathrm f}<f(j)+1$ was reached below $v_i$, since
valuation dropouts precede unit decreases at $v_i$. If the auction
reaches $v_i$, \Cref{lem:cap-band} gives
$x_j+b_j/v_i<y_j^{\mathrm f}+1$. If it ends earlier, the weak
bound at the terminal price becomes strict when the denominator
increases to $v_i$, since $b_j>0$.
\end{proof}

For $\lambda>0$, let
\[
\Pi_j(\lambda)\coloneqq\int_{(\lambda,\infty)}(c-\lambda)\,dx_j(c)
\]
be the total excess payment buyer~$j$ pays on purchases above $\lambda$.
Subtracting $\lambda$ times the quantity bought after $\lambda$
from the corresponding payments gives
\begin{align}\label{eq:premium-identity}
b_j(\lambda)+\lambda\bigl(x_j(\lambda)-y_j^{\mathrm f}\bigr)
=b_j^{\mathrm f}+\lambda z_j^{\mathrm f}+\Pi_j(\lambda).
\end{align}
Taking $\lambda=v_i$ and using \Cref{lem:local} gives
$b_j^{\mathrm f}<v_i(1-z_j^{\mathrm f})$, since $\Pi_j(v_i)\ge0$.

\begin{lemma}\label[lemma]{lem:exit-accounting}
If $\lambda>0$ and $x_j(c)>y_j^{\mathrm f}$ for every $c>\lambda$,
then
$\sum_{i\ne j}b_i^{\mathrm f}[1-\lambda/v_i]_+
\le\Pi_j(\lambda)$.
\end{lemma}

\begin{proof}
The hypothesis implies $x_j^{\mathrm f}>y_j^{\mathrm f}$, hence
$z_j^{\mathrm f}>0$. Fix $i\ne j$ with $b_i^{\mathrm f}>0$ and
$v_i>\lambda$. By \Cref{lem:final-divisible}, $i$ drops out at
$v_i$ while $j$ is active. Just before this dropout,
$x_j>y_j^{\mathrm f}\ge\mu_j(\bm d)$. Stability and
\Cref{lem:budget-equality} therefore give
$\bm b(A\setminus j)=v_i r$.

Removing $i$ lowers this sum by $b_i^{\mathrm f}$ and leaves $r$
unchanged. By \eqref{eq:clinch}, buyer~$j$ purchases at least
$b_i^{\mathrm f}/v_i$ units at price $v_i$, contributing at least
$b_i^{\mathrm f}(1-\lambda/v_i)$ to $\Pi_j(\lambda)$.
Summing over the dropouts proves the bound, including ties because
the auction processes them one at a time.
\end{proof}

\subsection{Choice of the Comparison Price}

\begin{lemma}[{\citet{SomaYoshida2018},
proof of Lemma~2}]\label[lemma]{lem:matching}
Let $\bm y,\bm y'\in\cBZ(f)$. For each $i$, form
$[y_i-y_i']_+$ left copies and $[y_i'-y_i]_+$ right copies.
The copies admit a perfect matching in which every matched pair
$(i,j)$ satisfies $\bm y-\unit_i+\unit_j\in\cBZ(f)$.
\end{lemma}

\begin{proof}[Proof of \Cref{lem:support-price}]
Fix $\bm y'\in\cBZ(f)$ and apply \Cref{lem:matching} to $\yf$
and $\bm y'$. Call a buyer who gains integer units a recipient.
For each recipient~$j$, let $\alpha_j$ be the lowest valuation
among buyers matched to her gained units.

Choose the comparison price by following the auction up to $\rho$.
For a recipient~$j$, the two relevant events are that the price
reaches $\alpha_j$ and that her quantity exceeds $y_j^{\mathrm f}$.
Take the infimum of the prices at which both conditions hold for
some recipient, with $\rho$ as an upper bound:
\begin{equation}\label{eq:choose-lambda}
\lambda\coloneqq\inf\bigl(
\{\rho\}\cup
\{c\ge0\mid \alpha_j\le c,\ x_j(c)>y_j^{\mathrm f}
                 \text{ for some recipient }j\}\bigr).
\end{equation}
Thus, $0<\lambda\le\rho$.

Now fix any $\bm z'\ge0$. From each buyer's payment bound subtract
$\lambda(z_i'-z_i^{\mathrm f})$, the change in her divisible
quantity valued at the comparison price. We bound the resulting
amount by
\begin{align*}
\min\{b_i^{\mathrm f},v_i\Delta x_i\}
-\lambda(z_i'-z_i^{\mathrm f})
\le {}&
\min\{v_i,\lambda\}(y_i'-y_i^{\mathrm f})
+b_i^{\mathrm f}[1-\lambda/v_i]_+.
\end{align*}
Indeed, bound the minimum on the left by the weighted average with
weight $\min\{v_i,\lambda\}/v_i$ on the valuation term.
The coefficient of $z_i'$ after subtraction is nonpositive, and
the coefficient of $z_i^{\mathrm f}$ vanishes whenever
$z_i^{\mathrm f}>0$, since then $v_i\ge\rho\ge\lambda$.

For every recipient~$j$ with $\alpha_j\le\lambda$ and
$\alpha_j<\min\{v_j,\rho\}$, we use a sharper bound.
If $\alpha_j<\lambda$, monotonicity of $x_j(c)$ and the choice
of $\lambda$ imply that $j$ held at most $y_j^{\mathrm f}$ units
before $\lambda$. She therefore bought at least $z_j^{\mathrm f}$
units at prices at or above $\lambda$. It follows that
$\Pi_j(\alpha_j)-\Pi_j(\lambda)
\ge(\lambda-\alpha_j)z_j^{\mathrm f}$,
which also holds when $\alpha_j=\lambda$.
Apply \Cref{lem:local} and \eqref{eq:premium-identity} to a matched
exchange whose donor has valuation $\alpha_j$. Using the budget
bound and $z_j'\ge0$ gives
\begin{align*}
\min\{b_j^{\mathrm f},v_j\Delta x_j\}
-\lambda(z_j'-z_j^{\mathrm f})+\Pi_j(\lambda)
\le b_j^{\mathrm f}+\alpha_j z_j^{\mathrm f}+\Pi_j(\alpha_j)
<\alpha_j.
\end{align*}

To sum these bounds, value each lost integer unit at its donor's
valuation capped at $\lambda$. For a recipient who uses the first
bound, $\alpha_j\ge\min\{v_j,\lambda\}$. Each gained unit is
therefore covered by its matched loss. For a recipient who uses
the sharper bound, one matched loss has capped value $\alpha_j$.
That loss alone strictly covers her adjusted payment together
with $\Pi_j(\lambda)$. Any other losses matched to her have
nonnegative capped value.

It remains to cover the budget terms
$b_i^{\mathrm f}[1-\lambda/v_i]_+$ in the first bound.
If $\lambda=\rho$, they all vanish. If $\lambda<\rho$, there is
a recipient~$j$ with $\alpha_j\le\lambda$ and
$x_j(c)>y_j^{\mathrm f}$ for every $c>\lambda$: there are finitely
many recipients, and each $x_j(c)$ is nondecreasing.
In particular, $z_j^{\mathrm f}>0$ and $v_j\ge\rho>\lambda$.
This recipient uses the sharper bound, and \Cref{lem:exit-accounting}
shows that her excess payment $\Pi_j(\lambda)$ covers the budget
terms for all other buyers.

Summing over buyers and using $\bm z^{\mathrm f}(N)=m$ gives
$\sum_i\min\{b_i^{\mathrm f},v_i\Delta x_i\}
-\lambda\bigl(\bm z'(N)-m\bigr)\le0$. 
If equality holds, no recipient uses the sharper bound.
Hence $\lambda=\rho$, and all budget terms in the first bounds
vanish. Summing these bounds and applying the matching yields 
$0\le
\sum_i\min\{v_i,\rho\}(y_i'-y_i^{\mathrm f})
\le0$, 
which proves the claim.
\end{proof}

\section{Proof of \Cref{thm:eligibility-impossible}}
\label[appendix]{app:eligibility-impossible}

Consider three buyers, one indivisible item, and one divisible unit.
Buyers~1 and~2 have unrestricted access to the divisible good, while
buyer~3 can receive at most $\eta$ units, where $0\le\eta<1$.
Put $\kappa\coloneqq1-\eta$ and set the budgets to
$\bm B=(2,1+\kappa/2,2)$. The divisible rank $g_\eta$ is $1$ on every
set containing buyer~1 or~2, with $g_\eta(\{3\})=\eta$ and
$g_\eta(\varnothing)=0$. The case $\eta=0$ gives
\Cref{thm:eligibility-impossible}. For $\eta>0$, all three buyers
have access to the divisible good.

\begin{theorem}\label[theorem]{pc:thm:main}
For every $0\le\eta<1$, no deterministic mechanism for this instance
with budget-feasible, nonnegative output payments satisfies IC, IR,
and PO for all valuations.
\end{theorem}

\begin{proof}
Suppose such a mechanism exists. Fix $(v_1,v_3)=(a,1)$ with
$1/2<a<1$, and let $t$ be buyer~2's report. We derive a necessary
condition for each possible holder of the item, then use these
conditions to show that raising $a$ can reduce buyer~1's quantity.

With other reports fixed, IC makes quantity nondecreasing in a
buyer's report and gives the same payment at reports yielding the
same quantity~\citep[Lemma~2]{Myerson1981}. At a jump between two constant quantities, IC on
either side implies that the payment increase equals the quantity
increase times the report at which the jump occurs. PO requires
full allocation, and Pareto comparisons allow negative payments. 
We first consider reports below $1$ to derive payment restrictions
at higher reports through IC.

\emph{Reports below $1$.}
Whenever buyer~3 does not hold the indivisible item, IR gives
$p_3\le z_3\le\eta<1$. With budget $2$, she can buy the item from
any holder whose valuation is below $1$ at that holder's valuation
and strictly gain. Thus, buyer~1 never holds the item and buyer~3
holds it whenever~$t<1$.

At such a report, buyer~3 also receives her full divisible allowance
$\eta$. Otherwise, $p_3\le1+z_3<2$, and she can buy some divisible
goods from a lower-valued buyer. For $t\ne a$, PO assigns the remaining $\kappa$ units to the higher-valued buyer, who retains budget by IR. 
Buyer~2 receives nothing and pays zero
for $t<a$. Her quantity increases by $\kappa$ at $a$, giving her quantity
and payment $(\kappa,a\kappa)$ for $a<t<1$.

\emph{Reports above $1$.}
For $t>1$, either buyer~3 holds the item (A) or buyer~2 holds it
(B). Write
\[
\gamma\coloneqq B_2-a\kappa-\eta=\kappa(3/2-a),
\qquad 0<\gamma<\kappa.
\]

In (A), buyer~2 receives the entire divisible unit whenever $t<B_2$:
she has the highest valuation and retains budget since
$p_2\le tx_2\le t<B_2$. In (B), she receives at least one unit
because she holds the item. Thus, $x_2\ge1$ just above $1$ and,
by monotonicity, at every higher report. If (A) occurs at $t$,
feasibility gives $x_2\le1$, so her quantity is exactly $1$
throughout $(1,t]$. In particular, whenever (A) occurs at some
$t>1$, buyer~2 receives the entire divisible unit, buyer~1 receives
nothing and pays zero, and buyer~3 holds only the item.
The increase from $\kappa$ to $1$ at report $1$
costs $\eta$. Her payment in (A) is therefore $a\kappa+\eta$, leaving
unspent budget $\gamma$.

In (B), let $z$ be buyer~2's divisible quantity. Compared with a
report just below $1$, her quantity increases by $1+z-\kappa=\eta+z$.
IC for types approaching $1$ from below gives
$p_2-a\kappa\ge\eta+z$. Hence, $z\le\gamma<\kappa$. Buyer~1 holds at least
$\kappa-z>0$ divisible units. If $z_3<\eta$, buyer~3 could buy some
of these units at $a$, since $p_3\le z_3\le \eta<1$. Thus, $z_3=\eta$,
and buyer~1 holds exactly $\kappa-z$.

\emph{Two Pareto comparisons.}
In (A), give the item to buyer~2 and assign divisible quantities
$(\kappa-\gamma/t,\gamma/t,\eta)$. Charge buyer~2 her remaining budget
$\gamma$ and buyer~1 the value $a(\kappa-\gamma/t)$ of her gain, which
she can afford from her initial payment of zero. Refund $\kappa$ to
buyer~3. These payment changes preserve each buyer's utility.
PO requires that they do not increase revenue, giving
$\gamma+a(\kappa-\gamma/t)\le\kappa$.
Since $\gamma-(1-a)\kappa=\kappa/2$, outcome~(A) requires
\begin{equation}\label{pc:eq:A}
t\le 2a\gamma/\kappa=a(3-2a).
\end{equation}

In (B), give the item to buyer~3 and the entire divisible unit to
buyer~2. Buyer~3 gains $\kappa$ units and can pay $\kappa$, since
$p_3+\kappa\le\eta+\kappa=1$. The refunds needed to preserve the utilities
of buyers~1 and~2 total $a(\kappa-z)+tz$. If these refunds were less
than $\kappa$, the seller could keep the difference. Thus, (B) requires
\begin{equation}\label{pc:eq:B}
(t-a)z\ge(1-a)\kappa.
\end{equation}

\emph{Contradiction.}
Fix $(v_2,v_3)=(17/16,1)$. When buyer~1 reports $a=11/20$,
(A) would require $17/16\le209/200$ by \eqref{pc:eq:A}.
Thus, (B) occurs. Since $z\le\gamma=19\kappa/20$, buyer~1 receives
at least $\kappa/20>0$ divisible units.
Now raise her report to $a=3/4$. In (B), the budget bound gives
$z\le3\kappa/4$, whereas \eqref{pc:eq:B} requires $(5/16)z\ge\kappa/4$.
These inequalities are incompatible because $15\kappa/64<\kappa/4$.
Thus, (A) occurs and buyer~1 receives nothing. Raising her report
has reduced her quantity, contradicting~IC.
\end{proof}

\end{document}